\documentclass[10pt,compsoc]{IEEEtran}

\makeatletter \def\mdseries@tt{m} \makeatother 

\usepackage{makecell}
\usepackage{tcolorbox}
\usepackage{indentfirst}
\usepackage{booktabs}
\usepackage{graphicx}
\graphicspath{./figures}
\usepackage[singlespacing]{setspace}
\usepackage{threeparttable}
\usepackage{multirow}
\usepackage{subcaption}
\usepackage[utf8]{inputenc}
\usepackage[english]{babel}
\usepackage{amsthm}
\usepackage{changepage} 
\usepackage{mathtools}
\usepackage{textgreek}
\usepackage[compact]{titlesec}         
\usepackage{amsmath}
\usepackage{tikz}
\usepackage{orcidlink}
\usetikzlibrary{shapes,positioning,arrows,calc}

\DeclareFontFamily{OT1}{pzc}{}
\DeclareFontShape{OT1}{pzc}{m}{it}{<-> s * [1.1] pzcmi7t}{}
\DeclareMathAlphabet{\mathpzc}{OT1}{pzc}{m}{it}
\theoremstyle{definition}
\newtheorem{definition}{Definition}
\DeclareMathAlphabet\mathbfcal{OMS}{cmsy}{b}{n}
\usepackage{centernot}
\usepackage{enumitem}
\makeatletter
\def\algbackskip{\hskip-\ALG@thistlm}
\usepackage{xcolor}
\usepackage[ruled, vlined, linesnumbered]{algorithm2e}
\newcommand{\nonl}{\renewcommand{\nl}{\let\nl\oldnl}}
\SetKwInput{KwInput}{Input}                
\SetKwInput{KwOutput}{Output} 
\SetAlFnt{\small\sffamily}

\SetCommentSty{mycommfont}
\usepackage{amsmath, listings, amsthm, amssymb, proof, xspace}
\usepackage{hyperref}
\hypersetup{colorlinks,allcolors=black}
\theoremstyle{remark}

\newcommand{\added}[1]{\textcolor{black}{{#1}}}
\newcommand{\minor}[1]{\textcolor{black}{{#1}}}
\lstdefinelanguage{ocl1}{
  keywords={init,Message, Pre, Post, Client,Range,Invariant,Constraints},
  keywordstyle=\color{purple}\ttfamily\bfseries,
  keywords=[2]{not, and, or, state, transition, implies, component,dbg,receipt,:,|,;},
  keywordstyle=[2]\color{black}\ttfamily\bfseries,
  sensitive=false,
  comment=[l]{//},
  morecomment=[s]{/*}{*/},
  commentstyle=\color{blue}\ttfamily,
  stringstyle=\color{red}\ttfamily,
  morestring=[b]',
  morestring=[b]"
}

\lstdefinelanguage{antlr}{
  keywords={grammar, rule,ExecRule,EOF},
  keywordstyle=\color{Purple}\ttfamily\bfseries\footnotesize,
  keywords=[2]{script:, execRule:,when:, where:, statements:, scriptStatement:,interactiveStatement:,umlrtCmd:,random:,dbgCommands:,|},
  keywordstyle=[2]\color{black}\ttfamily\bfseries\footnotesize,
  otherkeywords = {-,|},
  morekeywords = [3]{-},
  morekeywords = [4]{|},
  keywordstyle = [3]{\color{blue}}\footnotesize,
  keywordstyle = [4]{\color{blue}}\footnotesize,
  identifierstyle=\color{black}\footnotesize,
  sensitive=true,
  comment=[l]{//},
  morecomment=[s]{/*}{*/},
  commentstyle=\color{green}\ttfamily,
  stringstyle=\color{blue}\ttfamily,
  morestring=[b]',
  morestring=[b]",
  alsodigit={:}
}

\lstdefinelanguage{z3}{
	sensitive=true,
	alsoletter={\-},
	comment=[l]{;},
	keywords=[1]{
apply, assert, assert-soft, check-sat, check-sat-using, compute-interpolant,
declare-const, declare-datatypes, declare-fun, declare-map, declare-rel,
declare-sort, declare-tactic, define-sort, display, echo, eval, exit,
fixedpoint-pop, fixedpoint-push, get-assertions, get-assignment, get-info, get-
interpolant, get-model, get-option, get-proof, get-unsat-core, get-user-tactics,
get-value, help, help-tactic, labels, maximize, minimize, pop, push, query,
reset, rule, set-info, set-logic, set-option, simplify
	},
	morekeywords=[2]{
check-sat-using, declare-var, declare-rel, rule, query, set-predicate-
representation, maximize, minimize, assert-soft, assert-weighted, compute-
interpolant
	},
}

\author{Amirhossein~Zolfagharian~\orcidlink{0000-0002-2411-7938}, Manel~Abdellatif~\orcidlink{0000-0002-8647-1676}, ~Lionel~C.~Briand~\orcidlink{0000-0002-1393-1010},
Mojtaba~Bagherzadeh~\orcidlink{0000-0002-0253-671X} and~Ramesh~S
\IEEEcompsocitemizethanks{\IEEEcompsocthanksitem Amirhossein Zolfagharian is affiliated with the Department
of Electrical and Computer Engineering, University of Ottawa, Ottawa, Canada.\protect\\
E-mail: A.zlf@uottawa.ca
\IEEEcompsocthanksitem Manel Abdellatif is with the Software and Information Technology Engineering Department, École de Technologie Supérieure, Montreal, Canada. She contributed to this work mainly during her postdoctoral fellowship at the School of EECS, University of Ottawa, Ottawa, Canada.\protect\\
E-mail: Manel.abdellatif@etsmtl.ca
\IEEEcompsocthanksitem Lionel Briand is with the Department
of Electrical and Computer Engineering, University of Ottawa, Ottawa, Canada, and also with the SnT Centre for Security, Reliability and Trust, University of Luxembourg, Luxembourg.\protect\\
E-mail: Lbriand@uottawa.ca
\IEEEcompsocthanksitem Mojtaba Bagherzadeh is with Cisco, Ottawa, Canada. He contributed to this work mainly during his postdoctoral fellowship at the School of EECS, University of Ottawa, Ottawa, Canada.\protect\\
E-mail: Mbagherz@cisco.com
\IEEEcompsocthanksitem Ramesh S is with the Department of Research and Development, General
Motors, Warren, MI, USA.\protect\\
E-mail: Ramesh.s@gm.com
}

\thanks{Manuscript received June 15, 2022;  First revision January 13, 2023; Second revision April 06, 2023; Accepted April 15, 2023}}

\begin{document}

\title{A Search-Based Testing Approach for\\ Deep Reinforcement Learning Agents}

\IEEEtitleabstractindextext{%

\begin{abstract}
Deep Reinforcement Learning (DRL) algorithms have been increasingly employed during the last decade to solve various decision-making problems such as autonomous driving, trading decisions, and robotics. However, these algorithms have faced great challenges when deployed in safety-critical environments since they often exhibit erroneous behaviors that can lead to potentially critical errors. One of the ways to assess the safety of DRL agents is to test them to detect possible faults leading to critical failures during their execution. This raises the question of how we can efficiently test DRL policies to ensure their correctness and adherence to safety requirements.
Most existing works on testing DRL agents use adversarial attacks that perturb states or actions of the agent. However, such attacks often lead to unrealistic states of the environment. Furthermore, their main goal is to test the robustness of DRL agents rather than testing the compliance of the agents' policies with respect to requirements. Due to the huge state space of DRL environments, the high cost of test execution, and the black-box nature of DRL algorithms, exhaustive testing of DRL agents is impossible.

In this paper, we propose a Search-based Testing Approach of Reinforcement Learning Agents (STARLA) to test the policy of a DRL agent by effectively searching for failing executions of the agent within a limited testing budget. We rely on machine learning models and a dedicated genetic algorithm to narrow the search toward faulty episodes (i.e., sequences of states and actions produced by the DRL agent). \added{We apply STARLA on Deep-Q-Learning agents trained on two different RL problems widely used as benchmarks and show that STARLA
 significantly outperforms Random Testing by detecting more faults related to the agent's policy.} We also investigate how to extract rules that characterize faulty episodes of the DRL agent using our search results. Such rules can be used to understand the conditions under which the agent fails and thus assess the risks of deploying it.

\end{abstract}

}

\maketitle
\ifCLASSOPTIONcompsoc
\IEEEraisesectionheading{\section{Introduction}\label{Sec:Introduction}}
\else
\section{Introduction}
\label{Sec:Introduction}
\fi

\IEEEPARstart{R}{einforcement~Learning} (RL) algorithms have seen tremendous research advances in recent years, both from a theoretical standpoint and in their applications to solve real-world problems. Reinforcement Learning ~\cite{wiering2012reinforcement} trains an agent to make a sequence of decisions to reach a final goal and is therefore a technique gaining increased interest in many application contexts, such as autonomous driving and robotics. Deep Reinforcement Learning (DRL) techniques~\cite{sewak2019deep,lei2021deep,vanhasselt2015deep}, a branch of deep learning where RL policies are learned using Deep Neural Networks (DNNs), have gained attention in recent years. However, like DNN components, their application in production environments requires effective and systematic testing, especially when used in safety-critical applications. For instance, deploying a reinforcement learning agent in autonomous driving systems raises major safety concerns as we must pay attention not only to the extent to which the agent's objectives are met but also to damage avoidance~\cite{dulac2019challenges}.  

One of the ways to assess the safety of DRL agents is to test them to detect possible faults leading to critical failures during their execution. By definition, a fault in DRL-based systems corresponds to a problem in the RL policy that may lead to the agent's failure during execution.  
Since DRL techniques use DNNs, they inherit the advantages and drawbacks of such models, making their testing challenging and time-consuming. Furthermore, DRL-based systems are based on a Markov Decision Process (MDP)~\cite{10.5555/528623} that makes them stateful. They embed several components including the agent, the environment, and the ML-based policy network. Testing a stateful system that consists of several components is by itself a challenging problem. It becomes even more challenging when ML components and the probabilistic nature of the real-world environments are considered.

Furthermore, in a DRL context, two types of faults can be defined: functional and reward faults. The former happens when an RL agent takes an action that leads to an unsafe state (e.g., a driverless car does not stop at a stop sign). The latter occurs when an agent does not reach the desired reward (e.g., when a driverless car reaches its destination an hour late). Functional and reward faults are often in tension, as we can obtain a high reward while observing a functional fault.  
For example, an unsafe action can help the agent reach the desired state faster (e.g., when not stopping at stop signs, the driverless car may reach its destination sooner). This makes the detection of such types of faults challenging, especially when the agent is well-trained and failures are rare. We should note that we focus our analysis on functional faults since they are more critical. Thus, the detection of reward faults is left for future work and is out of the scope of this paper.

There are three types of testing approaches for deep learning systems that depend on the required access levels to the system under test: white-box, black-box, and data-box testing~\cite{trujillo2020does,lin2020black}. 
White-box testing requires access to the internals of the DL systems and its training dataset. Black-box testing does not require access to these elements and considers the DL model as a black box. Data-box testing requires access only to the training dataset.  

Prior testing approaches for DL systems (including DNNs) have focused on black-box and white-box testing~\cite{10041782,tian2018deeptest,nikanjam2022faults,xie2019deephunter,ilahi2021challenges,Huang2017AdversarialAO,pan2021improving}, depending on the required access level to the system under test. However, limited work has been done on testing DRL-based systems in general and using data-box testing methods in particular~\cite{nikanjam2022faults,ilahi2021challenges,pan2021improving,tan2020robustifying}. Relying on such types of testing is practically important, as testers often do not have full access to the internals of RL-based systems but do have access to the training dataset of the RL agent~\cite{byun2021black}.

Most existing works on testing DRL agents are based on adversarial attacks that aim to perturb states of the DRL environment~\cite{lin2019tactics}. However, adversarial attacks lead to unrealistic states and episodes, and their main objective is to test the RL agents' robustness rather than test the agents' functionality (e.g., functional safety). In addition, a white-box testing approach for DRL agents has been proposed that focuses on fault localization in the source code of DRL-based systems~\cite{nikanjam2022faults}. However, this testing approach requires full access to the internals of the DRL model, which are often not available to testers, especially when the DRL model is proprietary or provided by a third party. Also, localizing and fixing faults in the DRL source code do not prevent agent failures due to imperfect policies and the probabilistic nature of the RL environment. Furthermore, because of the huge state space, the high cost of test execution, and the black-box nature of DNN models (policy networks), exhaustive testing of DRL agents is impossible.

In this paper, we focus on testing the policies of DRL-based systems using a data-box testing approach, and thus address the needs of many practical situations.  
We propose STARLA, a Search-based Testing Approach for Reinforcement Learning Agents, that is focused on testing the agent's policy by searching for faulty episodes as effectively as possible. An episode is a sequence of states and actions that results from executing the RL agent. To create these episodes, we leverage evolutionary testing methods and rely on a dedicated genetic algorithm to identify and generate functional-faulty episodes~\cite{holland1992genetic}. We rely on state abstraction techniques~\cite{akrour2018regularizing,pmlr-v80-abel18a} to group similar states of the agent and significantly reduce the state space. We also make use of ML models to predict faults in episodes and guide the search toward faulty episodes. \added{We applied our testing approach on two Deep-Q-Learning agents trained for the widely known \textit{Cart-Pole} and \textit{Mountain Car} problems in the OpenAI Gym environment~\cite{Gymlibrary}.} We show that our testing approach outperforms Random Testing, as we find significantly more faults.

Overall, the main contributions of our paper are as follows:

\begin{itemize}
    \item We propose STARLA,  a data-box search-based approach to test DRL agents' policies by detecting functional-faulty episodes.

    \item We propose a highly accurate machine learning-based classification of RL episodes to predict functional-faulty episodes, which we use to improve the guidance of the search. This is based in part on defining and applying the notion of abstract state to increase learnability. 

    \item \added{We applied STARLA on two well-known RL problems. We show that STARLA outperforms the Random Testing of DRL agents as we detect significantly more faults when considering the same testing budget (i.e., the same number of the generated episodes).}
    
    \item We provide online\footnote{https://github.com/amirhosseinzlf/STARLA} a prototype tool for our search-based testing approach as well as all the needed data and configurations to replicate our experiments and results. 
\end{itemize}

The remainder of the paper is structured as follows.
Section~\ref{Sec:Background} presents the required background and establishes the context of our research. Section~\ref{Sec:Problem.Definition} describes our research problem. Section~\ref{Sec:Approach} presents our testing approach. Section~\ref{Sec:Evaluation} reports our empirical evaluation and results. 
Section~\ref{Sec:Discussitons} discusses the practical implications of our results. 
Section~\ref{Sec:Threats} analyzes the threats to validity. 
Finally, sections~\ref{Sec:RW} and~\ref{Sec:Conclusion} contrast our work with related work and conclude the paper, respectively.

\section{Background}
\label{Sec:Background}

Reinforcement Learning trains a model or an agent to make a sequence of decisions to reach a final goal. It is therefore a technique gaining increased interest in complex autonomous systems. RL uses trial and error to explore the environment by selecting an action from a set of possible actions. The actions are selected to optimize the obtained reward. In the following, we will describe an example application of RL in autonomous vehicles.  

\subsection{Definitions}

To formally define our testing framework, we rely on a running example and several key concepts that are introduced in the following sections.

\renewcommand\lstlistingname{$\line(1,0){240}$ Algorithm}

\textit{\textbf{{A running example.}}}
 Assuming an Autonomous Vehicle (AV) cruising on a highway, an RL agent (i.e., AV) receives observations from an RGB camera (placed in front of the car) and attempts to maximize its reward during the highway driving task, which is calculated based on the vehicle's speed. The agent is given one negative/positive reward per time step when the vehicle's speed is below/above 60 MPH (miles per hour). Available actions are turning right, turning left, going straight, and no-action. The cruising continues until one of the termination criteria is met: (1) the time budget of 10 seconds is consumed, or (2) a collision has occurred.

\begin{definition}\textit{({RL Agent Behavior.}})
\label{def-MDP} The behavior of an RL agent can be captured as a Markov Decision Process~\cite{10.5555/528623} $\langle S, A, T, R, \gamma \rangle$ where $S$ and $A$ denote a set of possible states and actions accordingly, $T: S \times A \times S \xrightarrow[]{} [0,1]$ refers to the transitions function, such that $T(s',a,s)$ determines 
the probability of reaching state $s'$ by performing action $a$ in state $s$, $R: S \times A \xrightarrow[]{} [0, R_{max}]$ is a reward function that determines the immediate reward for a pair of an action and a state, and $\gamma \in [0,1]$ is the discount factor indicating the difference of short-term and long-term reward~\cite{sutton2018reinforcement}.

The solution of an MDP is a policy $\pi : S \xrightarrow[]{} A$ that denotes the selected action given the state. The agent starts from the initial state ($s_0 \in S$) at time step $t=0$ and then, at each time step ($t_i$, $i\geq 0$), it takes an action ($a_i \in A$) according to the policy $\pi$ that results in moving to a new state $s_{i+1}$. Also, $r_i$ refers to the reward corresponding to the action $a_i$ and state $s_i$ that is obtained at the end of the step~$t_i$. Note that there may not be a reward at each step, which in that case is considered to be zero. Finally, $\sum^i{r_i}$ refers to the accumulative reward until step~$t_i$.

\end{definition}

\begin{definition}\textit{({Episodes.}})
\label{def-episode}
An episode $e$ is a finite sequence of pairs of states and actions, i.e., $[ (s_j,a_j) | s_j \in S, a_j \in A, 0 \le j \le n, n \in \mathbb{N} ]$, where the state of the first pair is an initial state, and the state of the last pair is an end state. An end state is, by definition, a state in which the agent can take no more action. The accumulative reward of episode $e$ is $\sum^{|e|}{r}$, where $|e|$ denotes the length of the episode. We refer to the accumulated reward of episode $e$ with $r'_e$. 
A valid episode is an episode where each state is reachable from the initial state with respect to the transition function presented in definition~\ref{def-MDP}. Moreover, the episode is executable (i.e., consistent with the policy of an agent) if starting from the same initial state and in each state, the selected action of the agent is consistent with the action in the episode that we want to execute.

\end{definition}

\begin{definition}\textit{({Faulty state.}})
\label{def-faulty-state}
A faulty state is a state in which one of the defined requirements (e.g., the autonomous vehicle must not hit obstacles) does not hold, regardless of the accumulated reward in that state. A faulty state is often an end state. In the context of the running example, a faulty state is a state where a collision occurs.

\end{definition}

\begin{definition}\textit{({Faulty Episode.}})
\label{def-faulty-episode}
We define two types of faulty episodes:
\begin{itemize}
    \item \textit{\textbf{Functional fault:}} If an episode $e$ contains a faulty state, it is considered as a faulty episode of type functional. A functional fault may lead to an unsafe situation in the context of safety-critical systems (e.g., hitting an obstacle in our running example).
    \item \textit{\textbf{Reward fault}:} If the accumulative reward of episode $e$ is less than a predefined threshold ($r'_e \leq \tau$), it is considered a faulty episode of type reward (i.e., the agent failed to reach the expected reward in the episode). Intuitively, regarding our running example, if we assume a reward fault threshold of $\tau = 100$, then each episode with a reward below 100 is considered to contain a reward fault. In our running example, this occurs when the AV agent drives at 25 MPH all the time. As we mentioned earlier, the detection of this type of fault is out of the scope of this paper and is left for future work.

\end{itemize}

\end{definition}

\subsection{State Abstraction}
\label{subsec:State abstraction}
State abstraction is a means to reduce the size of the state space by clustering similar states to reduce the complexity of the investigated problem~\cite{akrour2018regularizing,pmlr-v80-abel18a}. State abstraction can be defined as a mapping from an original state $s \in S$ to an abstract state $s^\phi \in S^\phi$
\begin{equation}
    \phi : S \xrightarrow{} S^\phi
\end{equation}

where the abstract state space is often much smaller than the original state space. Generally, there are three different classes of abstraction methods in the RL context~\cite{jiang2018notes,li2006towards}:
\begin{enumerate}
    \item $\pi^*$-irrelevance abstraction: $s_1$ and $s_2$ are in the same abstraction state $\phi(s_1)= \phi(s_2)$, if $\pi^*(s_1)=\pi^*(s_2)$,  where $\pi^*$ represents the optimal policy.
    \item $Q^*$-irrelevance abstraction: $\phi(s_1)= \phi(s_2)$ if for all available actions $a\in A$,  $Q^*(s_1,a)=Q^*(s_2,a)$, where $Q^*(s,a)$ is the optimal state-action function that returns the maximum expected reward from state $s$ up to the final state when selecting action $a$ in state $s$.
    \item Model-irrelevance abstraction: $\phi(s_1)= \phi(s_2)$ if for any action $a\in A$ and any abstract state $s^\phi \in S_\phi$,  $R(s_1,a)=R(s_2,a)$ and the transition dynamics of the environment are also similar, meaning that $\sum_{s' \in \phi^{-1}( s^\phi)} {T(s',a,s_1)} = \sum_{s' \in \phi^{-1}( s^\phi)}{T(s',a,s_2)}$ where $T(s',a,s)$ returns the probability of going to state $s'$ from state $s$ performing action $a$, as defined in definition~\ref{def-MDP}.
\end{enumerate}

As we are testing RL agents in our work, we use the $Q^*$-irrelevance abstraction method in this study because it represents the agent's perception. We also choose this abstraction method because it is more precise than $\pi^*$-irrelevance. Indeed, $\pi^*$-irrelevance only relies on the predicted action (i.e., the action with the highest $Q^*$-value) to compare two different states, which makes it coarse. In contrast, $Q^*$-irrelevance relies on $Q^*$-values for all possible actions.

To clarify, assume two different states in the real world for which our trained agent has the same Q-values. Given the objective to test the agent, it is logical to assume these states to be similar, as the agent has learned to predict identical state-action values for both states (i.e., the agent perceives both states to be the same). 

Further, abstraction methods can be strict or approximate. Strict abstraction methods use a strict equality condition when comparing the states of state-action pairs, as presented above. Although they are more precise, they bring limited benefits in terms of state space reduction. In contrast, more lenient abstraction methods can significantly reduce the state space, but they may yield inadequate precision.
Approximate abstractions relax the equality condition in strict abstraction methods to achieve a balance between state space reduction and precision. For example, instead of $Q^*(s_1,a)=Q^*(s_2,a)$, approximate abstraction methods use the condition  $|Q^*(s_1,a)-Q^*(s_2,a)|<\epsilon$, where $\epsilon$ is a parameter to control the trade-off between abstraction precision and state space reduction.

Another important property is transitivity, as transitive abstractions use a transitive predicate. For example, assume that two states,  $s_1$ and $s_2$, are similar based on an abstraction predicate and the same is true for $s_2$ and $s_3$. Then, we should be able to conclude that $s_1$ and $s_3$ are similar. Transitive abstractions are efficient to compute and preserve the near-optimal behavior of RL agents~\cite{pmlr-v80-abel18a}. Moreover, this property helps to create abstract states more effectively. 

Considering the properties that we explained previously, we use the following abstraction predicate $\phi_d$ that is transitive and approximates the $Q^*$-irrelevance abstraction:

\begin{equation}
\small
   \phi_d(s_1)=\phi_d(s_2) \equiv \forall a \in A : \bigg\lceil \frac{Q^*(s_1,a)}{d} \bigg\rceil = \bigg\lceil \frac{Q^*(s_2,a)}{d} \bigg\rceil
\end{equation}
 
where $d$ is a control parameter (abstraction level) that can squeeze more states together when increasing and thus reduce the state space significantly. Intuitively, this method discretizes the $Q^*$-values with buckets of size $d$.

\section{Problem Definition}
\label{Sec:Problem.Definition}

In this paper, we propose a systematic and automated approach to test a DRL agent. In other words, considering a limited testing budget, we aim to exercise the agent in a way that results in detecting faulty episodes, if possible. This requires finding faulty episodes in a large space of possible episodes while satisfying a given testing budget, defined as the number of executed episodes.

\subsection{RL Agent Testing Challenges}
\label{subsec:TestingChallenges}

Since DRL techniques use DNNs, they inherit the advantages and drawbacks of DNNs, making their testing challenging and time-consuming~\cite{dnnt2018arXiv180304792S, deepcon8802786,sun2018concolic,Ma_2018}. In addition, RL techniques raise specific challenges for testing:
\begin{itemize}

\item \textbf{Functional faults.}  The detection of functional faults in DRL systems is challenging because relying only on the agent's reward is not always sufficient to detect such faults. Indeed, an episode with a functional fault can reach a high reward. 
For example, by not stopping at stop signs, the car may reach its destination sooner and get a higher reward if the reward is defined based on the arrival time. Even if we consider a penalty for unsafe actions, we can still have an acceptable reward for functional-faulty episodes. Relying only on the agent's reward makes it challenging to identify functional faults.

\item \textbf{State-based testing with uncertainty.} Most traditional ML models, including DNNs, are stateless. However, DRL techniques are based on an MDP that makes them stateful and more difficult to test. Also, an RL agent's output is the result of an interaction between the environment (possibly consisting of several components, including ML components) and the agent. Testing a system with several components and many states is by itself a challenging problem. Accounting for ML components and the probabilistic nature of real-world environments makes such testing even more difficult~\cite{dulac2021challenges,trujillo2020does}.
    
\item \textbf{Cost of test execution.} According to the previous discussion, testing an RL agent requires the execution of test cases by either relying on a simulator or by replaying the captured logs of real systems. The latter is often limited since recording a sufficient number of logs that can exhaustively include the real system's behavior is impossible, especially in the context of safety-critical systems, for which logs of unsafe states cannot (easily) be captured. Thus, using a simulator for testing DRL agents, specifically in the context of safety-critical domains, is often inevitable. Despite significant progress made in simulation technology, high-fidelity simulators often require high-computational resources. Thus, testing DRL agents tends to be computationally expensive~\cite{ilahi2021challenges,dulac2021challenges,chaffre2020sim}.

\item \textbf{Focus on adversarial attacks.} Most existing works on testing DRL agents use adversarial attacks that are focused on perturbing states~\cite{Huang2017AdversarialAO}. However, such attacks lead to unrealistic states and episodes~\cite{ghamizi2020search}. The main goal of such attacks is to test the robustness of RL policies rather than the agents' functionality~\cite{carlini2017towards,ilahi2021challenges,zhang2020robust}. 

\end{itemize}

The exhaustive testing of DRL agents is impossible due to the large state space,  the black-box nature of DNN models (policy networks), and the high cost of test execution. To address these challenges, we propose a dedicated search-based testing approach for RL agents that aims to generate as many diverse faulty episodes as possible. To create the corresponding test cases, we leverage meta-heuristics and most particularly genetic algorithms that we tailor to the specific RL context.

\subsection{Assumptions}

In this work, we focus on testing RL agents with discrete actions and a deterministic policy interacting with a stochastic environment. A discrete action setting reduces the complexity of the problem in defining genetic search operators, as we will see in the following sections. It also reduces the space of possible episodes. Moreover, assuming a deterministic policy and stochastic environment is realistic because in many application domains (specifically in safety-critical domains), randomized actions are not acceptable and environments tend to be complex~\cite{dulacarnold2019challenges}. 
\added{We further assume that we have neither binary nor noisy rewards~\cite{zhang2020adaptive}~\cite{rakhsha2021reward} (i.e, where an adversary manipulates the reward to mislead the agent) since the reward function should provide guidance in our search process.} We build our work on model-free RL algorithms since they are more popular in practice and have been extensively researched~\cite{swazinna2022comparing,OpenAi}.
 
\section{Approach}
\label{Sec:Approach}

Genetic Algorithms (GA) are evolutionary search techniques that imitate the process of evolution to solve optimization problems, especially when traditional approaches are ineffective or inefficient~\cite{alsmadi2010using}. In this research, as for many other test automation problems, we use genetic algorithms to test RL agents. This is accomplished by analyzing the episodes performed by an RL agent to generate and execute new episodes with high fault probabilities from a large search space.

 \begin{figure*}[ht]
    \centering
    \includegraphics[width=\textwidth]{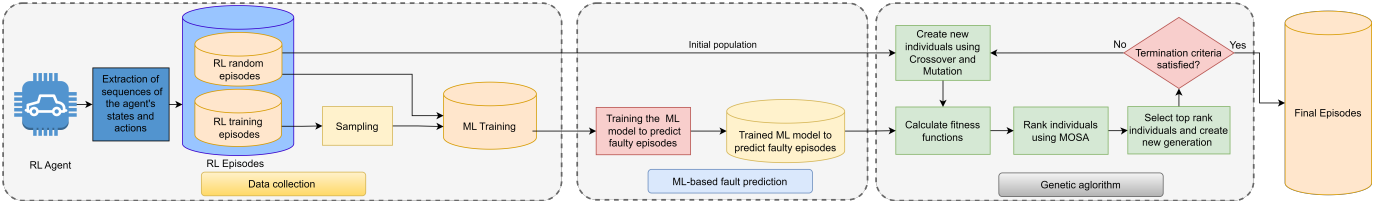}
    \caption{Overview of STARLA}
    \label{fig:Approach}
  \end{figure*}

\subsection{Reformulation as a Search Problem}
\label{Sec:Reformulation}

We are dealing with a high number of episodes represented as sequences of pairs (Definition~\ref{def-episode}), which are executed to test an RL agent. To properly translate the process into a search problem using a genetic algorithm, we need to define the following elements.
\begin{itemize}
    \item \textbf{Individuals.} Individuals consist of a set of elements called genes. These genes connect together and form an individual that is a solution. Here, individuals are episodes complying with Definition~\ref{def-episode}, which should ideally have a high probability of failure. Naturally, each gene is represented by a pair of state and action.

    \item \textbf{Initial population.} A set of individuals forms a population. In our context, a population is a set of episodes. However, it is imperative for the search to select a diverse set of individuals to use as the initial population. The sampling process is detailed in section~\ref{subsec:initial-population}.
    
    \item \textbf{Operators.} Genetic operators include crossover, mutation, and selection~\cite{whitley1994genetic}, which are used to create the next generation of episodes. In the crossover, we use two episodes as input and create a new offspring that hopefully has a higher fault probability. We use the current population as the input and select an episode for the crossover (using tournament selection). The selection of such an episode is in relation to its fitness. We then select a crossover point randomly, search for a matching episode, and join it with the selected episode.
    
    Mutation is an operator that adds diversity to our solutions. An episode is selected using the tournament selection again. Then, one pair is randomly selected as the mutation point, which is altered according to a defined policy that is detailed in section~\ref{subsec:mutation}.
    
    Selection is the last operator used in each generation. It combines the episode from the last generation with the newly created episode in a way that does not eliminate good solutions from previous generations.
    More detailed explanations are provided in section~\ref{sec:Selection}.

    \item \textbf{Fitness function.} The fitness function should indirectly capture how likely an episode is to be faulty. To that end, we define a multi-objective fitness~\cite{murata1995moga} function that we use to compare episodes and select the fittest ones. As further explained in section~\ref{Sec:Fitness}, we consider different ways to indirectly capture an episode's faultiness: (1) the reward loss, (2) the predicted probability of observing functional faults in an episode based on machine learning, and (3) the certainty level for the actions taken in an episode. 
    
    \item \textbf{Termination criteria.} This determines when the search process should end. Different termination criteria can be used, such as the number of generations or iterations, the search time, and convergence in population fitness. For the latter, the search stops when there is no improvement above a certain threshold over a number of newly generated episodes.
    
\end{itemize}

\subsection{Overview of the Approach}
\label{Sec:Overview}

As depicted in Figure~\ref{fig:Approach}, the main objective of STARLA is to generate and find episodes with high fault probabilities to assess whether an RL agent can be safely deployed. 

To apply a genetic algorithm to our problem, we first need to sample a diverse subset of episodes to use as the initial population. In the next step, we use dedicated genetic operators to create offspring to form the new population. Finally, using a selection method, we transfer individuals from the old population to the new one while preserving the diversity of the latter. For each fitness function, we have a threshold value, and a fitness function is satisfied if an episode has a fitness value beyond that threshold. We repeat this process until all fitness functions are satisfied, or until the maximum number of generations is reached.

\begin{algorithm}[t!]

\DontPrintSemicolon
  \KwInput{A set of episodes ($P$) as the initial population, solution archive $\alpha$}  

  \KwOutput{The updated archive $\alpha$ containing faulty episodes}
   
    $gen \xleftarrow{} 0$ \;
    $\alpha \xleftarrow{} \emptyset$ \;
    \While{$ FitSatisfied = False  \And  gen \leq g $}{
    $Fitness= Fit(P)$\;
    $P_{new} \xleftarrow{} \emptyset$\;
    $rand = Random(0,1) $\;
    \If{$rand < c$}{
    $t_{1},t_{2} \xleftarrow{} Cross(P)$\;
    $P_{new} \xleftarrow{} P_{new} \cup \{t_{1}, t_{2}\} $\;
    }
    $t_m \xleftarrow{} Mut(P\cup \{t_{1}, t_{2}\} , m )$\;
    $P_{new} \xleftarrow{} P_{new} \cup t_m$\;
    $Fitness= Fit(P_{new})$\;
    $Update  (FitSatisfied)$\;
    $Update (\alpha)$ \;
    $P \xleftarrow{}  Select (P, P_{new})$\;
    $gen \xleftarrow{} gen +1$ \;
    }
    \KwRet{$\alpha$}
\caption{High-level genetic algorithm}
\label{Algorithm-overview}
\end{algorithm}

Algorithm~\ref{Algorithm-overview} shows a high-level algorithm for the process previously described, based on a genetic algorithm. 
Assuming $P$ is the initial population, that is, a set of \textit{episodes} containing both faulty and non-faulty episodes, the algorithm starts an iterative process, taking the following actions at each generation until the termination criteria are met (lines 3-16). The search process is as follows: 
\begin{enumerate}
        \item We create a new empty population $P_{new}$ (line 5).
        \item We create offspring using crossover and mutation, and add newly created individuals to $P_{new}$ (lines 6-11).
        \item We calculate the fitness of the new population (line 12) and update the archive $\alpha$ and the condition $FitSatisfied$ capturing whether all fitness functions are satisfied (line 13). The archive contains all solutions that satisfy at least one of our three fitness functions (line 14).  
        \item If all fitness functions are satisfied, we stop the process and return the archive (i.e., the set of solutions that satisfy at least one fitness function). Otherwise, the population $P$ is updated using the selection function, and then we move to the next generation (lines 15-16). 
\end{enumerate}
Furthermore, in our genetic search algorithm, we set the crossover rate $c$ to 75\% and the mutation rate $m$ to the $\frac{1}{V}$, where $V$ is the length of the selected episodes for mutation based on the suggested parameters for genetic algorithms in the literature~\cite{MOSA7102604}.
In the following, we discuss each step of the search in detail.

\subsection{Initial Population}
\label{subsec:initial-population}

The initial population of our search problem is a set containing $|P|$ episodes.
We use random executions of the agent to build the initial population of the generic search. Consequently,  we initiate the environment with different initial states in which we randomly change the alterable parameters of the environment when available (e.g., changing the weather or time of the day in the running example, or changing the starting position of the car). We execute the RL agent starting from the randomly selected initial states and store the generated episodes as the initial population of the search. 

\subsection{Fitness Computations} 
\label{Sec:Fitness}

A fitness function quantitatively assesses the extent to which an individual fits the search objectives and is meant to effectively guide the search. Recall that our objective is to find faulty episodes that can exhibit functional faults. \added{We therefore define the following three fitness functions that complement one another to capture the extent to which an episode is close to being faulty. We consider (1) the agent reward as a fitness function, as it  guides the search toward low-reward episodes, which are more likely to lead to functional faults, if designed properly, as the reward may capture some of the agent's unsafe behavior; (2) the probability of functional faults, which complement the reward and can provide additional guidance toward functional faults if the estimation of such probabilities is accurate (see section \ref{subsec:TestingChallenges} for more explanation); and (3) the certainty level that guides the search toward episodes where the agent is highly uncertain about the selected action. }

\minor{Next, we define our fitness functions and illustrate their relevance through scenario examples (sections~\ref{Sec:Reward},~\ref{Sec:FunctionalFault}, and~\ref{Sec:CertaintyLevel}). Specifically, we explain how each fitness function contributes to guiding the search toward faulty episodes. We then explain in sections~\ref{Sec:ML},~\ref{subsec:Prepare training data ML},~\ref{subsec: State Abstraction for Training}, and~\ref{subsubSec:FeatureRepresnetation} the different steps of applying machine
learning to estimate the probability of functional faults,
which is one of our fitness functions. Finally, we explain our search objectives in section~\ref{Sec:Search.Algorithm}.  }

\subsubsection{Reward} 
\label{Sec:Reward}

The first fitness function in our search is meant to drive the search toward episodes with low reward. The reward fitness function of an episode is defined as follows.  
    \begin{equation}
    \label{eq:f1}
        f_1(e) = r'_e 
    \end{equation}
    where $r'_e$ is the accumulated reward of an episode $e$.
    
In the initial population, the rewards of selected episodes are known. Also, when genetic operators are applied, we calculate the reward of new individuals using the reward function according to Definition~\ref{def-MDP}.
The search aims to minimize the reward of episodes over generations to guide the search toward finding faulty episodes.

\minor{\textbf{\textit{Example:}} A well-defined reward function which is correlated to functional faults can provide valuable guidance toward the identification of such faults. For instance, in our running example, if the reward function incorporates a penalty based on the car's proximity to the border of the lane or other vehicles, minimizing the reward function would result in the car being driven dangerously close to other lanes, thereby increasing the risk of accidents.  Such low-reward episodes are of particular interest, as applying crossover and mutation (section~\ref{Sec:SearchOperators}) to them can reveal potential functional faults in the later generations. However, it is worth mentioning that designing a reward function that is capable of capturing all possible functional faults remains a challenging task in most real-world scenarios. We therefore rely on other complementary fitness functions (certainty level and the probability of functional faults) that we describe in the following sections to further guide the search toward functional faulty episodes.}

 \subsubsection{Probability of Functional Fault} 
\label{Sec:FunctionalFault}

The second fitness function captures the probability of an episode to contain a functional fault. Such probability is predicted using an ML model. The fitness function is defined as follows: 
    \begin{equation}
        f_2(e) = 1 - Pr_{f}[e]
    \end{equation}
    where $Pr_{f}[e]$ is the probability of having a functional fault in an episode $e \in E$, and $E$ is the space of all possible episodes. In the context of the running example, driving very close to obstacles has a higher probability of revealing a functional fault (high probability of collision) than an episode that maintains a safe distance from them. We therefore want the first episode to be favored by our search over the second one. 
    The ML model that we use takes an episode as input,  \added{uses the presence and absence of abstract states in the episode as features}, and returns the probability of functional fault for that episode. 
    A detailed explanation of the probability prediction method using ML is provided in section~\ref{Sec:ML}. 
   In the search process, instead of maximizing $Pr_{f}[e]$, we minimize its negation $f_2(e)$ to (1) have a consistent minimization problem across all fitness functions, and (2) guide the search toward finding episodes with a high probability of functional faults. 

\minor{\textbf{\textit{Example:}} In the context of our running example, it is possible to encounter a scenario where the speed of the vehicle is above 60 MPH (getting a positive reward at each time step) but excessively high, resulting in challenges to control the car, thus increasing the risk of collision. In such a case, using the reward fitness function is inadequate to guide the search toward such faulty episodes since they get high rewards. To address such situations, we can leverage the fitness function based on the probability of functional faults, which is predicted based on the presence of abstract states. With this approach, we can identify episodes that have a high probability of resulting in functional faults, even when the reward function cannot guide the search process effectively.}  
    
\subsubsection{Certainty Level} 
\label{Sec:CertaintyLevel}

This fitness function captures the level of certainty associated with the actions taken in each state within an episode. It is calculated as the average difference in each state-action pair between the probability of the chosen action, assigned according to the learned policy, and the second-highest probability assigned to an action~\cite{10.5555/3172077.3172414}.

A higher accumulated certainty level across the sequence of actions in an episode suggests that the agent is more confident overall about the selected actions. 
On the other hand, a lower accumulated certainty level can guide our search toward situations in which the agent is highly uncertain of the selected action.
Thus, it is relatively easier to lure the agent to take another action, which makes these episodes suitable for applying search operators.

The certainty level is calculated as shown in Equation~\ref{eq:f4}, where $e$ is the given episode, $|e|$ is its length, $a_i$ is the selected action in state $s_i$ (i.e., $a_i$ is the action with the highest selection probability), $A_i$ is the set of possible actions in state $s_i$, and $P_r(a_i|s_i)$ is the probability of selecting $a_i$ in state $s_i$. 
    \begin{equation}
    \label{eq:f4}
        f_3(e) = \frac{\sum_{i=1}^{|e|}{(P_r(a_i|s_i) - \max\limits_{a_j\in A_i \:\: \&\:\: j\neq i}P_r(a_j|s_i))}}{|e|}
    \end{equation}

In our search algorithm, we aim to minimize this fitness function to guide the search toward finding episodes with high uncertainty levels.

\minor{\textbf{\textit{Example:}} Suppose that in our running example, we have episodes where (1) the reward is high (i.e., driving with a speed above 60 MPH); (2) the probability of functional fault is low (i.e., the speed is not too high and the risk of collision is low); and (3) the uncertainty of the agent in selecting the optimal action is high. Such episodes are of particular interest, as they are promising candidates for our mutation operator (section~\ref{subsec:mutation}). Indeed, by applying a small realistic transformation to these episodes, we can easily change the optimal action and explore new search directions that have the potential to reveal functional faults. This fitness function is especially useful when the reward function and probability of functional fault cannot guide the search process effectively, and we need to explore different actions and states to identify potential functional faults. Incorporating the certainty level metric into our fitness functions can thus help us to identify such episodes more effectively. In other words, by targeting episodes with high uncertainty levels, we can increase our chances of discovering new functional faults and improve the overall performance of our search process.}

\subsubsection{Machine Learning for Estimating Probabilities of Functional Faults} 
\label{Sec:ML}

A machine learning algorithm is used to learn functional faults and estimate their probabilities in episodes without executing them.
This model is expected to take episodes as input and predict the probabilities of functional faults. The labels of each episode are functional-faulty or not faulty.
We choose \textit{Random Forest} as a candidate modeling technique because (1) it can scale to numerous features, and (2) its robustness to overfitting has been well studied in the literature~\cite{10.1023/A:1010933404324,friedman2001greedy}. \added{We also tried several other ML models to predict functional faults, such as \textit{K-Nearest Neighbor}, \textit{Support Vector Machine}, and \textit{Decision Trees}. However, \textit{Random Forest} led to the most accurate prediction model. Since this is not a crucial or central aspect of the work, we do not include the results of these experiments in the paper. }

\subsubsection{Preparation of Training Data}
\label{subsec:Prepare training data ML}
To build the above-mentioned machine learning model, the training data are collected from training episodes and random executions of the RL agent. More precisely, our ML training dataset contains both faulty and non-faulty episodes generated through the training and random executions of the agent. 

\noindent{\textbf{Episodes from RL training.}} We sample episodes from the agent's training phase to increase the diversity of the dataset. We also include such episodes in case we do not find enough faulty episodes based on random executions. Providing data with different types of episodes (i.e., functional-faulty and non-faulty) makes training the ML models possible. Since the training phase of the RL agents is exploratory, it contains a diverse set of faulty and non-faulty episodes, which helps learning and increases model accuracy.
One issue with sampling from the training episodes is that they may not be consistent with the final policy of the trained agent. 
The agent may execute a faulty episode during training because of (1) randomness in action selection, due to the exploratory nature of the training process, and (2) incomplete agent training. To alleviate this issue, when sampling to form the training dataset of the ML model, we give a higher selection probability to the episodes executed in the later stages of the training, since they are more likely to be consistent with the final behavior of the trained agent.

Assuming a sequence of $n$ episodes ([$E_i: 1\leqslant i \leqslant n$]) that are explored during the training of the RL agent, the probability of selecting episode $Ei$ ($P_r[E_i]$) is calculated as follows.

\begin{equation}
\label{eq:late-episode}
    P_r[E_i] = \frac{i}{\Sigma_{j=1}^{n}{j}}
\end{equation}

We thus give a higher selection probability to episodes executed in the later stage of the training phase of the agent ($P_r[E_1] < P_r[E_2] <...< P_r[E_n]$).

\noindent{\textbf{Episodes from random executions.} } To build the training dataset of our ML model, we also include episodes generated through random executions of the agent to further diversify the training dataset with episodes that are consistent with the final policy of the agent. In practice, we use the episodes of the initial population of the generic search since they have been already created with random executions of the RL agent (section~\ref{subsec:initial-population}), thus minimizing the number of simulations (and therefore the testing budget).

\subsubsection{State Abstraction for Training Data}\label{subsec: State Abstraction for Training}

After collecting the training episodes, we need to map each concrete state to its corresponding abstract state to reduce the state space and thus enable effective learning. Indeed, this is meant to facilitate the use of machine learning with more abstract features.
To do so, we rely on the transitive $Q^*$-irrelevance abstraction method, which was described in section~\ref{subsec:State abstraction}.

The state abstraction process is defined in Algorithm~\ref{Algorithm-abstract-classes}. The algorithm takes the concrete states as input and finds abstract states $s^{\phi_d} \in S_{^\phi}$ considering the abstraction function of $\phi_d$ where $d$ is the abstraction level. 
For each concrete state, we try to find the abstract state that corresponds to the concrete state by calculating the $Q^*$-values of all available actions, as described in section~\ref{subsec:State abstraction}. If a match with an abstract state of a previous concrete state that was already processed is found, we assign the abstract state to the concrete state. Otherwise, we create a new abstract state.

\begin{algorithm}[t!]
\DontPrintSemicolon
    \KwInput{Set of states $S$, abstraction level $d$ }
    \KwOutput{Abstract states $S^{\phi_d}$}
    $S^{\phi_d} \xleftarrow{} \emptyset$ \;
    $len \xleftarrow{} 0$\;
    \For{$s_i \in S$}{
    \If{$S^{\phi_d} = \emptyset$}
    {
    $len \xleftarrow{} len+1$\;
    append $s_i $ to $S^{\phi_d}_1$\;
    }
    $Found = False$
    
    \For {$j\:\: in \:\: range(1, len)$}
    {
    \If{$\phi(s_i)=S^{\phi_d}_j$}
    {append $s_i $ to $S^{\phi_d}_j$\;
     $Found = True$
    }
    }
    \If{Found = False}
    {
    $len \xleftarrow{} len+1$\;
    append $s_i$ to $S^{\phi_d}_{len}$\;
    
    }
    
    } 
    \KwRet{$S^{\phi_d}$} 
    \caption{High-level algorithm to create abstract states}
\label{Algorithm-abstract-classes}
\end{algorithm}

\subsubsection{Feature Representation: Presence and Absence of Abstract States}\label{subsubSec:FeatureRepresnetation}

To enable effective learning, each episode consists of state-action pairs, where the states are abstract states instead of concrete states. 
To train the ML model, we determine whether abstract states are present in episodes and use this information as features. As described in the following, each episode is encoded with a feature vector of binary values denoting the presence (1) or absence (0) of an abstract state $S^\phi_i$ in the episode and $n$ is the total number of abstract states.

\[ 
\begin{matrix}
& S^\phi_1  & S^\phi_2  & \cdots  &  S^\phi_i & \cdots & S^\phi_n\\
episode_i & 0 & 1& \cdots &0&\cdots&1
\end{matrix}
\]

The main advantage of this representation is that it is amenable to the training of standard machine learning classification models. Furthermore, we were able to significantly reduce the feature space by grouping similar concrete states through state abstraction, where the selected action of the agent is the same for all concrete states within one abstract state. As a result, considering $n$ different abstract states, the feature space of this representation is $2^n$. Note that we only consider the abstract states that have been observed in the training dataset of the ML model, which we expect to be rather complete. Further, in this feature representation, the order of the abstract states in the episodes is not accounted for, which might be a weakness if we are not able to predict functional faults as a result. Empirical results will tell whether the two above-mentioned potential problems materialize in practice.

\subsubsection{Multi-Objective Search}
\label{Sec:Search.Algorithm}

We need to minimize the above-mentioned fitness functions to achieve our goal, and this is therefore a multi-objective search problem.
More specifically, our multi-objective optimization problem can be formalized as follows:

\begin{equation}
\begin{split}
\min\limits_{x \in E} \:\:& F(x) =  (f_1(x),f_2(x),f_3(x))\\ 
\end{split}
\end{equation}

where E is the set of possible episodes in the search space, $F: E \xrightarrow{} \mathbf{R}^3$ consists of three real-value objective functions $f_1(x),f_2(x),f_3(x)$, and $\mathbf{R}^3$ is the objective space of our optimization problem. 

\subsection{Search Operators}
\label{Sec:SearchOperators}

We describe below three genetic operators. The first operator is crossover, which generates new offspring using slicing and joining high-fitness, selected individuals. 
The second operator is mutation, which introduces small changes in individuals to add diversity to the population, thus making the search more exploratory.
Finally, the selection operator determines which individuals survive to the next generation. We provide a detailed description of how we defined these operators.

\subsubsection{Crossover}
\label{Subsec:Crossover}

The crossover process is described in Algorithm~\ref{Algorithm-cross}. It uses the population as input and creates offspring as output. It begins by sampling an episode (line 2) with the $sample$ function. This function draws an episode using tournament selection~\cite{10.1162/evco.1996.4.2.113}. In a K-way tournament selection, K individuals are selected, and we run a tournament between the selected individuals where fitter individuals are more likely to be selected for reproduction.
Then, we randomly select a crossover point (line 3) using the uniform distribution.

After finding the crossover point, we must find a matching pair (line 4). We do so by considering individuals in the population containing the abstract state of the pair selected as a crossover point. The $search$ function tries to find a matching pair for the crossover point based on the $Q^*$-irrelevance abstraction method (section~\ref{subsec:State abstraction}). 
If no matching pair is found (line 5), we repeat the process from the beginning (lines 1-5). 
Otherwise, offspring are created on lines 6-9.  \added{Whether a match can be found for the crossover point highly depends on the abstraction level. Therefore, this can be controlled by changing the abstraction level to prevent bottlenecks.} 

\minor{The crossover process is illustrated in Figure~\ref{fig:Crossover}.} Let us assume that the selected parent is as follows:

\begin{equation}
\label{eq:random-Cpoint }
\begin{split}
Parent = [(s_{1},a_1),(s_2,a_2),...,&(s_{f-1},a_{f-1}),\boldsymbol{(s_{f},a_f)}\\
,(s_{f+1},a_{f+1}&) , ... ,(s_{m},a_m)]
\end{split}    
\end{equation}
where $(s_{f},a_f)$ is the pair selected as a crossover point.

The matching function tries to find an episode containing a pair that has a concrete state that belongs to the same abstract class as state $(s_{f})$  to ensure the validity of the new episode. Recall that all states in the same abstraction class are perceived to be the same by the RL agent. Also, since they have the same $Q^*$-values, their certainty level is the same.

\begin{equation}
\begin{split}
Match = [(s'_{1},a'_1),(s'_{2},a'_2),& ...,(s'_{v-1},a'_{v-1})\\
,\boldsymbol{(s'_{v},a'_{v})} , ... ,(s'_{n}&,a'_n)]
\end{split}    
\end{equation}

where $s'_{v}$ and $s_{f}$ result into the same abstract state based on our abstraction method (\minor{i.e., $\phi_d(s'_{v})=\phi_d(s_{f})$}). As a result, the selected actions are also the same.

The newly created offspring are:

\begin{equation}
\begin{split}
\mathit{Offspring_1} = [(s_1&,a_1),...,(s_{f-1},a_{f-1})\\
,(s'_{v},a'_{v}) , ... ,(s'_{n}&,a'_n) ]
\end{split}    
\end{equation}

\begin{equation}
\begin{split}
\mathit{Offspring_2} = [(s'_1&,a'_1),...,(s'_{v-1},a'_{v-1})\\
,(s_{f},a_{f}) , ... ,(s_{m}&,a_m)]
\end{split}    
\end{equation}

The first offspring contains the first part of the matching individual up to the crossover point with state $s_{f-1}$ and the second part is taken from the parent and vice versa for the second offspring. 

\begin{algorithm}[t!]
\DontPrintSemicolon
  \KwInput{Population $P$}
    \KwOutput{Offspring $B_1$ and $B_2$}
    \SetKwRepeat{Do}{do}{while}

     \Do{$match = \emptyset$}{ $parent \xleftarrow{} sample(P) $\;
    $l \xleftarrow{} CrossoverPoint(parent) $\;
    $match \xleftarrow{} search(P, parent[l])$\; }

    $B_1[0:l] \xleftarrow{} match[0:l]$\;
    
    $B_1[l:end] \xleftarrow{} parent[l:end]$\;
    $B_2[0:l] \xleftarrow{} parent[0:l]$\;
    $B_2[l:end] \xleftarrow{} match[l:end]$\;
    \KwRet{$B_1,B_2$}
\caption{High-Level Crossover Algorithm}
\label{Algorithm-cross}
\end{algorithm}

\minor{
\begin{figure}[ht]
    \centering
    \includegraphics[scale=0.47]{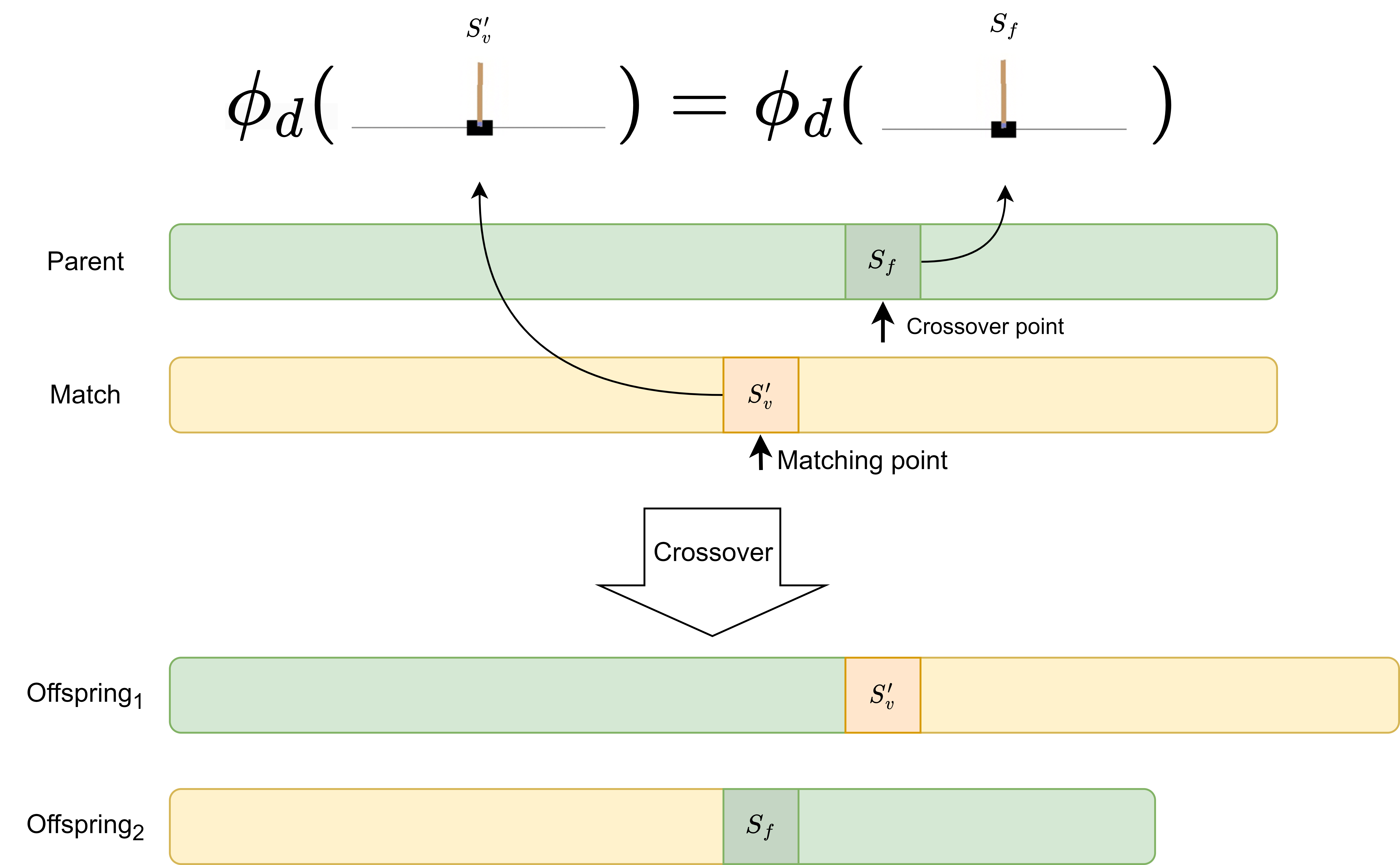}
    \caption{The crossover operator}
    \label{fig:Crossover}
\end{figure}
}

Based on the selected state abstraction method (section~\ref{subsec:State abstraction}), we create episodes that are more likely to be valid, though this is not guaranteed. 
Also, we may get inconsistent episodes (i.e., episodes that cannot be executed by the RL agent). Furthermore, due to the high simulation cost of the RL environment, we are not executing episodes after applying crossover during the search. The validity of the episodes in the final archive is therefore checked by executing the final high-fitness episodes. The execution process is described in detail in section~\ref{sec:execute}.

\subsubsection{Mutation}
\label{subsec:mutation}

The mutation operator starts by selecting an episode using a K-way tournament selection. 
Then a mutation point is randomly selected using the uniform distribution. 
To ensure the exploratory aspect of the mutation operator, we alter the state of the mutation point using some image transformation methods that are selected by considering the environment and the learning task to produce realistic and feasible states~\cite{deepx10.1145/3361566,tian2018deeptest}. 
These transformations are context-dependent. They should be realistic and representative of situations with imprecise sensors or actuators, as well as external factors that are not observable. In our running example, transformations matching such situations include changing the brightness and contrast of the image, adding tiny black rectangles to simulate dust on the camera lens, or changing the weather. Another example is the \textit{Cart-Pole} problem (section~\ref{subsec:Case}), where the task is to balance a pole on a moving cart. For this type of environment,  we rely on other transformations, such as slightly changing the position, the velocity, and the angle of the pole. 

After mutating the gene, we run the episode. Although executing episodes is computationally costly, mutation is infrequent and helps create valid and consistent episodes exploring unseen parts of the search space. Then, the updated episode is added to the population. Also, if we find any failure during the execution of the mutated episodes, we mark such episodes as failing and exhibiting a functional fault.

Assume that (1) the selected episode for mutation is $e_h$, (2) we select $(s_{c},a_{c})$ from $e_h$ as a candidate pair, and (3) the mutated/transformed state for $s_c$ is $s^t_{c}$. The mutated episode $e^m_h$ is then as follows:

\begin{equation}
\begin{split}
e_h = [(s_1,a_1),...,\boldsymbol{(s_{c},a_c)}&,(s_{c+1},a_{c+1}) ,\\
 ... ,(s_{m}&,a_m) ]
\end{split}    
\end{equation}

\begin{equation}
\begin{split}
e^m_h = [(s_1,a_1),...,\boldsymbol{(s^t_{c},a^t_{c})}, ...] 
\end{split}    
\end{equation}
where the states after the mutation point are determined from executing episode $e^m_h$.

\subsubsection{Selection} \label{sec:Selection}

\minor{Given that our search involves optimizing multiple fitness functions and the possibility of incorporating additional fitness functions to address different types of faults, it is imperative to employ a search algorithm that can handle more than three fitness functions. Additionally, it is worth noting that our focus is not on Pareto front optimization or the generation of a well-distributed set of solutions that capture the trade-off between objectives. Rather, we need a search algorithm that can optimize all the fitness functions concurrently and separately.}

 \minor{Considering these search requirements, we do not rely on traditional dominance-based algorithms like NSGA-II, particle swarm optimization (PSO)~\cite{wang2018particle}, or SPEA-II~\cite{4339452} because they show poor performance in problems with many objectives~\cite{8031325} and they target Pareto front optimization. Additionally, to build a generalizable and extensible testing approach, we opt for the Many Objective Sorting Algorithm (MOSA)~\cite{MOSA7102604} to select the best individuals that minimize our fitness functions. This is because we anticipate that, in many cases, we will need to consider more than three fitness functions, and MOSA is specifically tailored to our application context, software test automation. MOSA can indeed accommodate a large number of objectives and is better at generating diverse solutions to address all of the fitness functions. MOSA is therefore an optimal choice for our problem domain because no other search algorithm fulfills our requirements.
}

MOSA is a dominance-based multi-objective search algorithm based on NSGA-II~\cite{996017}. Although traditional dominance-based algorithms like NSGA-II and SPEA-II~\cite{4339452} show poor performance in problems with many objectives~\cite{8031325}, MOSA performs well even for a large number of objectives. It is widely used in the literature and tries to generate solutions that cover the fitness functions separately, instead of finding a well-distributed set of solutions (i.e., diverse trade-offs between fitness functions). More specifically, in our context, MOSA's purpose is to generate faulty episodes that separately satisfy at least one of the search objectives rather than to find episodes that capture diverse trade-offs among them. It drives the search toward the yet-uncovered search objectives and stores in an archive all solutions that satisfy at least one search objective.

MOSA works as follows. Similar to NSGA-II, it starts from an initial population and generates new offspring at each generation using genetic operators (i.e., mutation and crossover). We calculate the fitness value of each individual in the population based on the three fitness functions that we described in section~\ref{Sec:Fitness}. MOSA then uses a novel preference method to rank the non-dominant solutions. In this preference method, the best solutions according to each fitness function are rewarded with the $rank = 0 $, and the other solutions are ranked based on the traditional non-dominated sorting in NSGA-II. During the transition to a new population, we select the highest-ranked individuals using MOSA and add them to the new population without any changes. We also transfer a subset of the individuals with the highest fitness from the previous population to avoid losing the best solutions. Finally, an archive is used to store the best individuals for each individual fitness function.

\subsection{Execution of Final Results}
\label{sec:execute}

After completing the execution of the genetic algorithm, we obtain a population that contains episodes with high fault probability.   
We need to execute these final episodes to check their validity, their consistency with the policy of the agent, and whether they actually trigger failures. We assume that an episode is consistent if the RL agent can execute it. We retain failing episodes that are both valid and consistent.

During the execution process, we may observe deviations where the agent selects an action other than the action in the episode. To deal with such deviations, we replace the state observed by the agent with the corresponding state from the episode and observe the selected action.
For example, let us assume that we want to execute an episode $e'$ produced by STARLA where $e'=[ (s'_i,a'_i) | s'_i \in S, a'_i \in A, 0 \le i \le n, n \in \mathbb{N} ]$.
We set the state of the simulator to the initial state of the episodes $s'_0$. Then we use the states from the environment as input to the agent and we check the action selected by the agent. If during the execution, the agent selects an unexpected action $a_i \neq a'_i$ at state $s_i$, we replace $s_i$ with state $s'_i$ from episode $e'$ to drive the agent to select action $a'_i$.

If the action selected by the agent is not $a'_i$, we consider that episode $e'$ is invalid and we remove it from the final results.

Replacing states in this situation is acceptable since (1) we assume that the environment is stochastic, (2) states in episode $e'$ are real concrete states generated in the environment, and (3) we noticed that the states of the environment and in the episodes where deviations occur are very similar. The latter is likely due to the selection of the crossover point based on identical abstract states. Indeed, we observed that 94\% of the environment and episode states where deviations occur in the \textit{Cart-Pole} environment, which we will describe in detail in section~\ref{subsec:Cartpole}, have a cosine distance lower than 0.25.\added{ Similarly, we observed that 99\% of the deviations in the \textit{Mountain Car} environment (see section~\ref{subsec:MTC} for more details) have a cosine distance lower than 0.25.} Replacing similar states where deviations of the agent occur is therefore a sensible way to execute such episodes (if possible) because, in real-world environments, we may have incomplete or noisy observations due to imperfect sensors.

\section{Empirical Evaluation}
\label{Sec:Evaluation}

This section describes the empirical evaluation of our approach, including the research questions, the case study, the experiments, and the results. 

\subsection{Research Questions}
Our empirical evaluation is designed to answer the following research questions.

\subsubsection{RQ1. Do we find more faults than  Random Testing with the same testing budget?} 
We aim to study the effectiveness of our testing approach in terms of the number of detected faults compared to Random Testing. We want to compare the two approaches with the same testing budget, which is defined as the number of executed episodes during the testing phase. Given that the cost of real-world RL simulations can be high (e.g., autonomous driving systems), this is the main cost factor.

\subsubsection{RQ2. Can we rely on ML models to predict faulty episodes?}

In this research question, we want to investigate whether it is possible to predict faulty episodes using an ML classifier. We do not execute all episodes during the search; therefore, we want to use the probabilities of functional faults that are estimated by an ML classifier as a fitness function to guide our search toward finding faulty episodes. 

\subsubsection{RQ3. Can we learn accurate rules to characterize the faulty episodes of RL agents?}

One of the goals of testing an RL agent is to understand the conditions under which the agent fails. This can help developers assess the risks of deploying the RL agent and focus its retraining. Therefore, we aim to investigate the learning of interpretable rules that characterize faulty episodes from the final episodes that are executed once the search is complete. 

\subsection{Case Studies}\label{subsec:Case}

 \added{In our study, we consider two Deep-Q-Learning (DQN) agents on the \textit{Cart-Pole}\footnote{gymlibrary.dev/environments/classic\textunderscore control/cart\textunderscore pole/} balancing problem and \textit{Mountain Car} \footnote{gymlibrary.dev/environments/classic\textunderscore control/mountain\textunderscore car/}, both from the \textit{OpenAI Gym} environment.}

\added{We have chosen these RL case studies  because they are open source and widely used as  benchmark problems in the RL literature~\cite{nagendra2017comparison,aguilar2014stabilization,pattanaik2018robust}. We have also considered these benchmarks as they include a large number of concrete states. Furthermore, the simulations in such environments are fast enough to enable large-scale experimentation.}

 \subsubsection{Cart-Pole Balancing Problem}\label{subsec:Cartpole}
In the \textit{Cart-Pole} balancing problem, a pole is attached to a cart, which moves along a track. The movement of the cart is bidirectional and restricted to a horizontal axis with a defined range. The goal is to balance the pole by moving the cart left or right and changing its velocity.

\begin{figure}[ht]
    \centering
    \includegraphics[scale = 0.45]{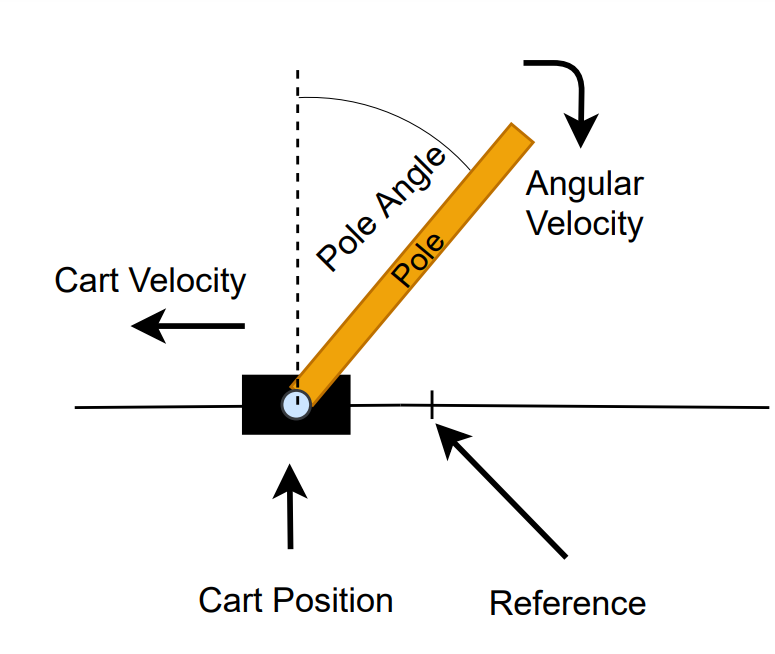}
    \caption{\textit{Cart-Pole} balancing problem}
    \label{fig:CartpoleExample}
\end{figure}

As depicted in Figure~\ref{fig:CartpoleExample}, the state of the agent is characterized by four variables: 
\begin{itemize}
    \item The position of the cart.
    \item The velocity of the cart.
    \item The angle of the pole.
    \item The angular velocity of the pole.
\end{itemize}

We provide a reward of +1 for each time step when the pole is still upright. The episodes end in three cases: (1) the cart is away from the center with a distance more than 2.4 units, (2) the pole's angle is more than 12 degrees from vertical, or (3) the pole remains upright during 200 time steps.
We define functional faults in the \textit{Cart-Pole} balancing problem as follows.

If, in a given episode, the cart moves away from the center by a distance above 2.4 units, regardless of the accumulated reward, we consider that there is a functional fault in that episode. \added{Note that termination based on the pole's angle is an expected behavior of the agent and thus a normal execution, whereas termination based on passing the borders of the track can cause damage and is therefore considered a safety violation.}

\begin{figure}[ht]
    \centering
    \includegraphics[scale=0.22]{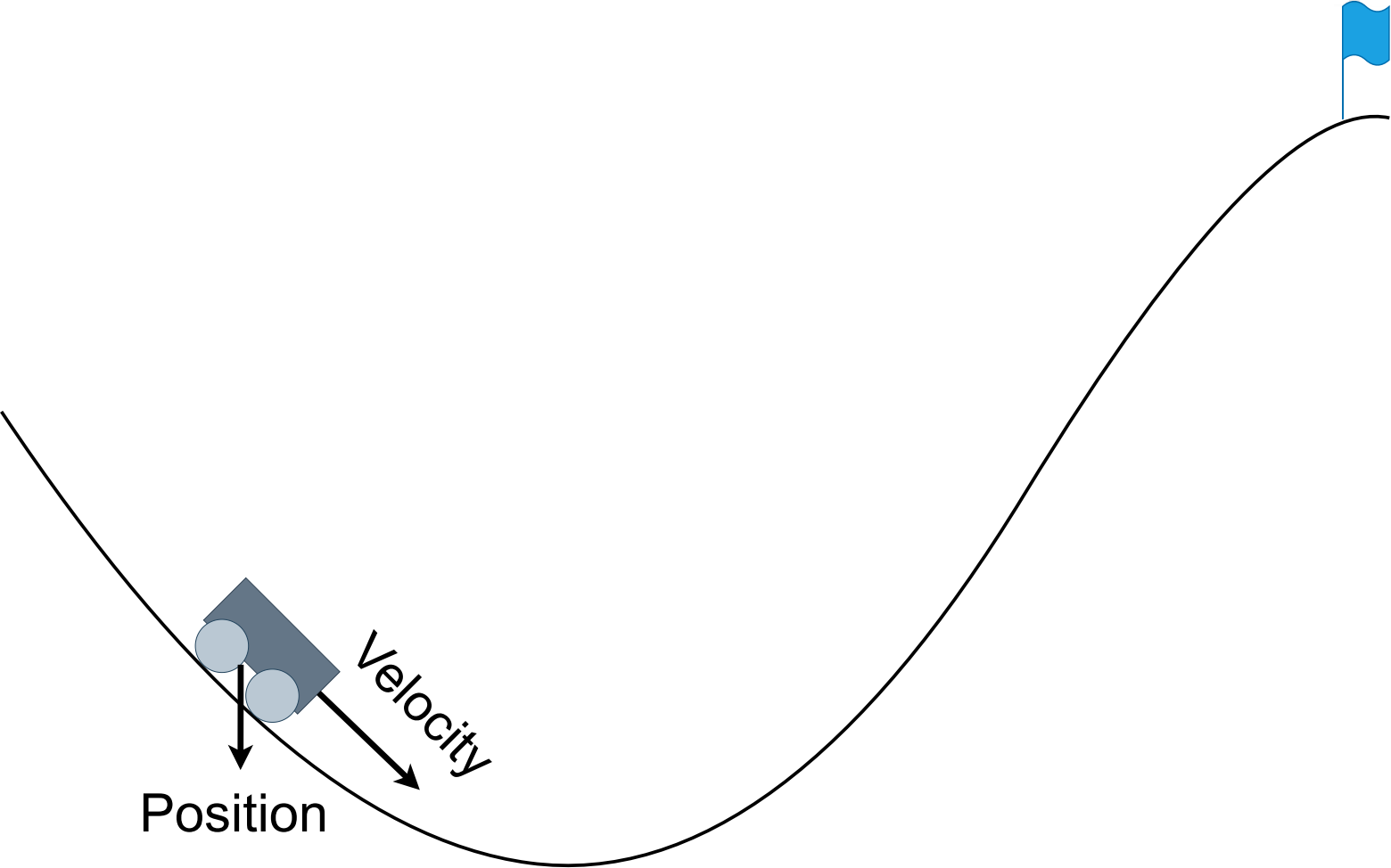}
    \caption{\textit{Mountain Car} problem}
    \label{fig:MountainCarExample}
\end{figure}

\added{\subsubsection{Mountain Car Problem}\label{subsec:MTC}
In the \textit{Mountain Car} problem, an under-powered car is located in a valley between two hills. 
The objective is to control the car and strategically use its momentum to reach the goal state on top of the right hill as soon as possible. The agent is penalized by -1 for each time step until termination. As illustrated in Figure~\ref{fig:MountainCarExample}, the state of the agent is defined based on (1) the location of the car along the x-axis, and (2) the velocity of the car.}

\added{There are three discrete actions that can be used to control the car:
\begin{itemize}
    \item Accelerate to the left.
    \item Accelerate to the right. 
    \item Do not accelerate.
\end{itemize}}

\added{Episodes end in three cases: (1) reaching the goal state, (2) crossing the left border, or (3) exceeding the limit of 200 time steps. In our custom version of the \textit{Mountain Car} (Figure~\ref{fig:MountainCarExample}), climbing the left hill is considered an unsafe situation. Consequently, reaching the leftmost position in the environment results in a termination with the lowest reward. Therefore, if in a given episode, the car crosses the left border of the environment,  we consider that there is a functional fault in that episode.}

\subsection{Implementation}
\label{Sec:Implementation}

\added{We used Google Colab and stable baselines~\cite{stable-baselines} to implement RL agents for both case studies (see section~\ref{subsec:Case}). Our RL agents are based on a DQN policy network~\cite{mnih2013playing} using standard setting of stable baselines (i.e., Double Q-learning~\cite{vanhasselt2015deep}, and dueling DQN~\cite{wang2016dueling}).}
Our \textit{Cart-Pole} RL agent has been trained for 50 000 time steps.
The average reward of the trained agent is 124 (which is also equal to the average length of the episodes). In general, the pole is upright over 124 time steps out of a maximum of 200. \added{The \textit{Mountain Car} agent has been trained for 90 000 time steps. The average reward is -125 and the average length of the episodes is 112.}

Finally, we execute the search algorithm for a maximum of 10 generations for both case studies. 
\added{The mean execution time of STARLA on Google Colab was 89 minutes for the \textit{Cart-Pole} problem and 65 minutes for the \textit{Mountain Car} problem.}

\subsection{Evaluation and Results}

\subsubsection{RQ1. Do we find more faults than Random Testing with the same testing budget?}

In this research question, we want to study STARLA's effectiveness in finding more faults than Random Testing when we consider the same testing budget \textit{B}, measured as the number of executed episodes.
To do so, we consider two practical testing scenarios: 
\begin{itemize}
    \item \textbf{Randomly executed episodes are available or inexpensive}: In the first scenario, we assume that we want to further test a DRL agent provided by a third-party organization. 
    We assume that both training episodes and some randomly executed episodes of the RL agent, used for testing the agent, are provided by the third party. Therefore, we can extract ML training data and an initial population from such episodes without using our testing budget. 
    
    We can also consider another situation where the RL agent is trained and tested using both a simulator and hardware in the loop~\cite{marco2017virtual}. Such two-stage learning of RL agents has been widely studied in the literature, where an agent is trained and tested on a simulator to ``\textit{warm-start}'' the learning on real hardware~\cite{marco2017virtual,kober2013reinforcement}. Since STARLA produces episodes with a high fault probability, we can use it to test the agent when executed on real hardware to further assess the reliability of the agent. In this situation, STARLA uses prior episodes that have been generated on the simulator to build the initial population and executes the newly generated episodes on the hardware. 
     
    In this case, randomly executed episodes using a simulator become relatively inexpensive. Therefore, only episodes that are executed with hardware in the loop and in the real environment are accounted for in the testing budget.  
    
    To summarize, when randomly executed episodes are available or inexpensive, the testing budget \textit{B} is equal to the sum of (1) the number of mutated episodes that have been executed during the search, and (2) the number of faulty episodes generated by STARLA that have been executed after the search.

    \item \textbf{Randomly executed episodes are generated with STARLA and should be accounted for in the testing budget:} In the second scenario, we assume that the agent is trained and then tested by the same organization using STARLA. Therefore, we have access to the training dataset but need to use part of our testing budget, using random executions, to generate the initial population.
    More precisely, the total testing budget in this scenario is equal to the sum of (1) the number of episodes in the initial population that have been generated through random executions of the agent; (2) the number of mutated episodes that have been executed during the search; and (3) the number of faulty episodes generated by STARLA that have been executed after the search. 

\end{itemize}
Because of randomness in our search approach and its significant execution time (section~\ref{Sec:Implementation}),\added{ for both case studies,} we re-executed the search algorithm 20 times and stored the generated episodes and the executed episodes with mutations at each run. We computed the mean number of generated functional-faulty episodes \textit{N} \added{(in \textit{Cart-Pole} \textit{$N_c$}=5313 and in \textit{Mountain Car} \textit{$N_m$} = 2809)} and the mean number of mutated executed episodes \textit{M} \added{(in \textit{Cart-Pole} case study\textit{$M_c$}=128 and in \textit{Mountain Car} \textit{$M_m=139$})} over the 20 runs. We analyzed the distribution of the total number of functional faults identified with STARLA over the 20 runs.
Then, we randomly selected (with replacement) 100 samples from the set of episodes that were generated with Random Testing. Each sample contained \textit{B} episodes to ensure that we had the same testing budget as in STARLA. 

In the \textit{Cart-Pole} case study, for the first scenario, \textit{B} is equal to 5441 (which corresponds to the mean number of generated faulty episodes and executed mutated episodes with STARLA over the 20 runs). On the other hand, for the second scenario, \textit{B} is equal to 6941 because, as previously explained, this testing budget accounts for the number of episodes in the initial population, which were generated with Random Testing (1500).
\added{Also, in the \textit{Mountain Car} case study, the mean number of generated faulty episodes and executed mutated episodes over the 20 runs is equal to 2948 (B = 2948 in scenario 1). For the second scenario, B is equal to 4448.}

\begin{figure}[ht]
    \centering
    \includegraphics[width=\columnwidth]{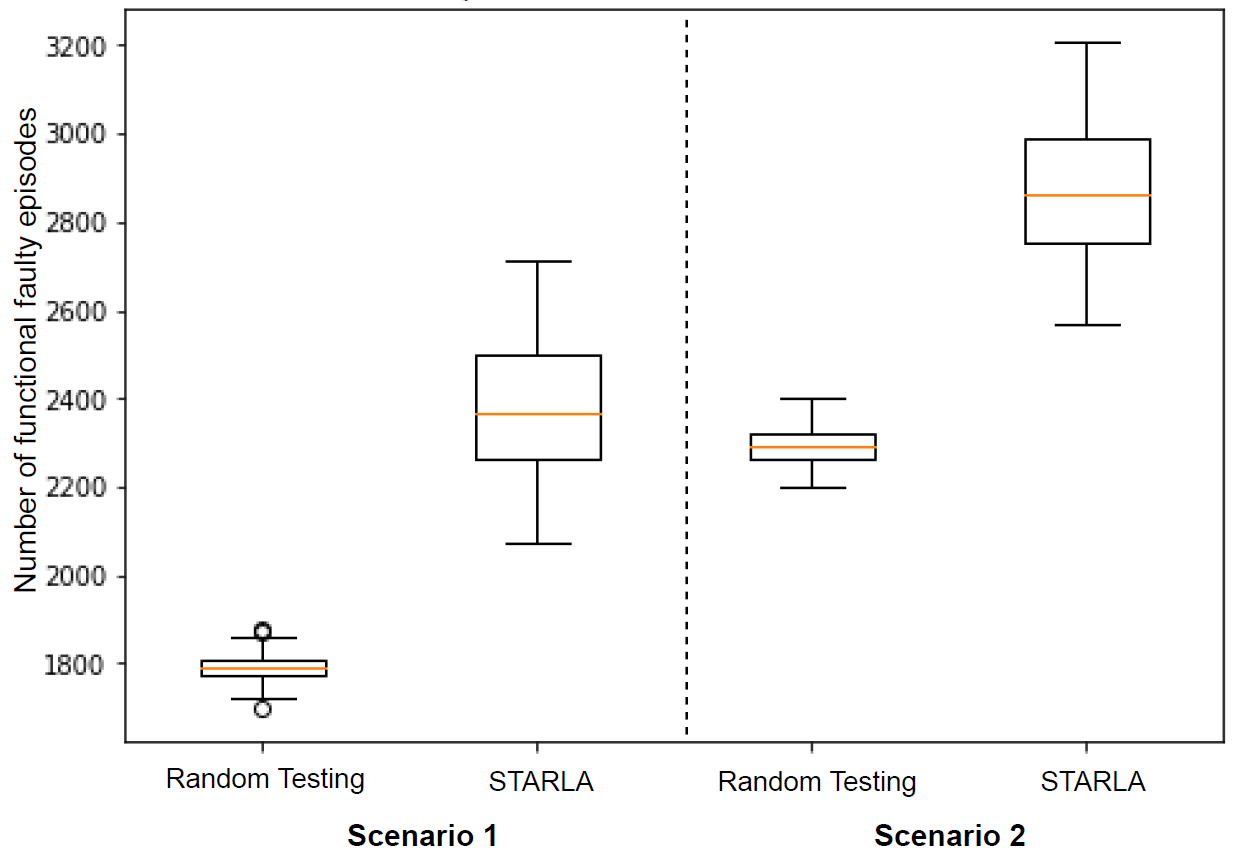}
    \caption{Number of functional-faulty episodes generated with STARLA compared to Random Testing in the \textit{Cart-Pole} problem}
    \label{fig:RQ1-boxplot-functional}
  \end{figure}

\begin{figure}[ht]
    \centering
    \includegraphics[width=\columnwidth]{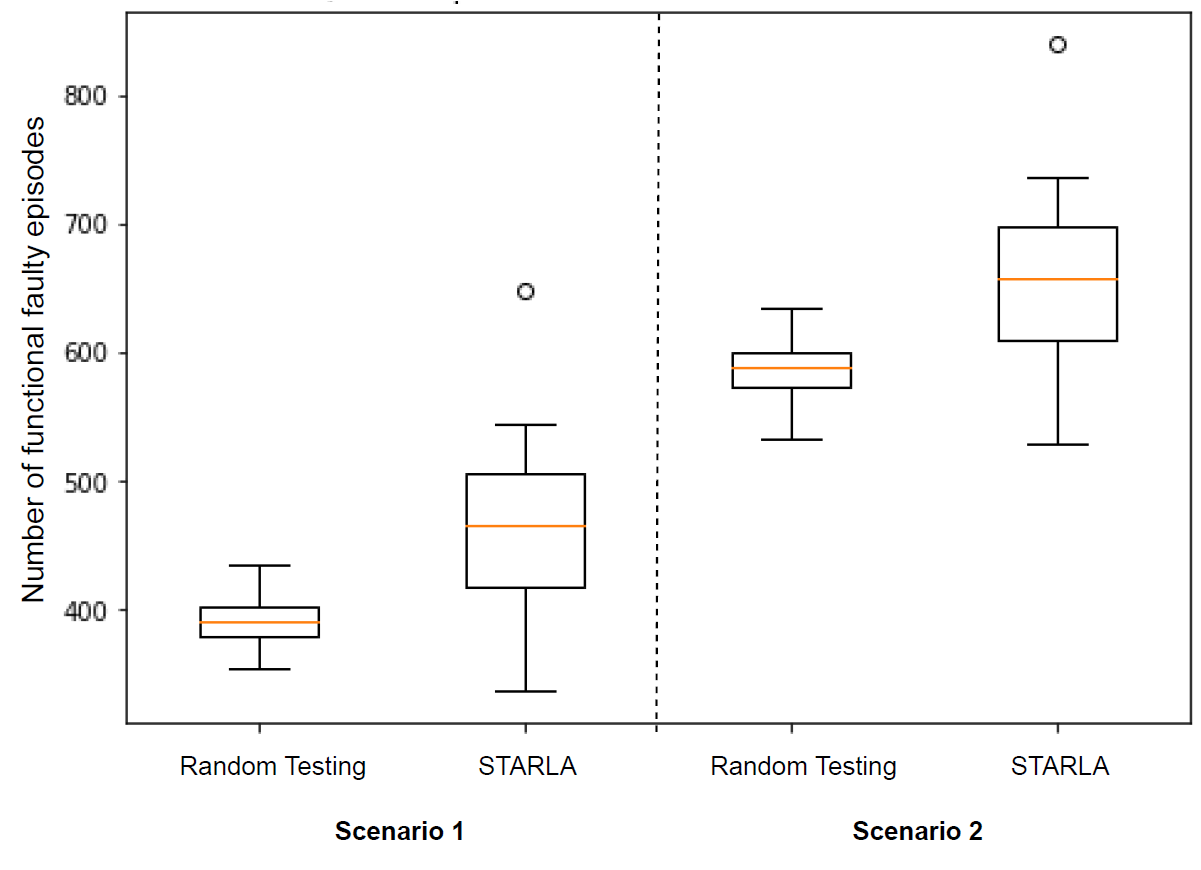}
    \caption{\added{Number of functional-faulty episodes generated with STARLA compared to Random Testing in the \textit{Mountain Car} problem}}
    \label{fig:RQ1_MC-boxplot-functional}
  \end{figure}

We analyzed the distribution of the identified faults in the two testing scenarios, compared it with STARLA, and reported the results. The results of the \textit{Cart-Pole} and \textit{Mountain Car} case studies are depicted in Figure~\ref{fig:RQ1-boxplot-functional} and Figure~\ref{fig:RQ1_MC-boxplot-functional}, respectively. We should note that in the first scenario, we only compute faults that are generated with the genetic search. We do not consider faults that are in the initial population because we assume that these episodes are provided to STARLA and are not included in the testing budget. In contrast, in the second scenario, we include in STARLA's final results the faulty episodes in the initial population as they are part of our testing approach and are included in the testing budget.   

\added{As we can see from the boxplots, our approach outperforms Random Testing in detecting faults in both scenarios. Indeed, in the first scenario, the average number of faulty episodes detected by STARLA is 2367, while an average of 1789 faulty episodes is detected with Random Testing in the \textit{Cart-Pole} case study.
 In the \textit{Mountain Car} case study, the average number of faulty episodes detected by STARLA is 466 while an average of 390 faulty episodes is detected with Random Testing (+19.5\%).}

\added{In the second scenario, where we consider a bigger testing budget, STARLA also outperforms Random Testing by identifying, on average, 2861 faulty episodes compared to 2291 identified by Random Testing (+24.9\%) in the \textit{Cart-Pole} case study, and 657 faulty episodes compared to 591 identified by Random Testing.} To assess the statistical significance of the average difference of the number of detected functional faults between STARLA and Random Testing, we use the non-parametric Mann-Whitney U-test~\cite{mcknight2010mann} and compute the corresponding p-value in both testing scenarios and case studies.

\added{In both scenarios for the two case studies, the p-values are far below 0.01, and thus show that our approach significantly outperforms Random Testing in detecting functional faults in DRL-based systems. }

\begin{tcolorbox}
\textbf{Answer to RQ1:} We find significantly more functional faults with STARLA than with Random Testing using the same testing budget. 
\end{tcolorbox}

\subsubsection{RQ2. Can we rely on ML models to predict faulty episodes?}
\label{RQ2}

We investigate the accuracy of the ML classifier to predict faulty episodes of the RL agent. We use \textit{Random Forest} to predict the probability of functional faults in a given episode.   
To build our training dataset in the \textit{Cart-Pole} case study, we sampled 2111 episodes, including 1500 episodes generated through random executions of the agent and 611 episodes from the agent's training phase (as described in section~\ref{subsec:initial-population}). Of these episodes, 733 correspond to functional faults while 1378 are non-functional faulty. 
\added{Further, the ML training dataset of the \textit{Mountain Car} case study is created by sampling 2260 episodes where (1) 1862 episodes are generated through random executions of the agent, and (2) 398 episodes are from the training phase of the \textit{Mountain Car} agent. Of these episodes, 456 are functional-faulty while 1804 are not.} 

\begin{table}[ht]
\centering

\begin{tabular}{|c|c|c|c|c|}
\hline
\textit{\textbf{d}} & \begin{tabular}[c]{@{}c@{}}\textbf{Abstract} \\ \textbf{states} \end{tabular} & \textbf{Accuracy} & \textbf{Precision} & \textbf{Recall} \\ \hline \hline
0.005 & 195073 & 63\% & 39\% & 63\% \\ \hline
0.01  & 146840 & 63\% & 39\% & 63\% \\ \hline
0.05  & 33574  & 73\% & 81\% & 73\% \\ \hline
0.1   & 15206  & 92\% & 92\% & 92\% \\ \hline
0.5   & 2269   & 95\% & 95\% & 95\% \\ \hline
1     & 1035   & 97\% & 97\% & 97\% \\ \hline
5     & 134    & 84\% & 84\% & 84\% \\ \hline
10    & 48     & 79\% & 81\% & 79\% \\ \hline
50    & 8      & 77\% & 78\% & 77\% \\ \hline
100   & 4      & 77\% & 78\% & 77\% \\ \hline
\end{tabular}

\caption{Prediction of \textit{functional faults} with \textit{Random Forest} in the \textit{Cart-Pole} case study. The first column represents the abstraction level \textit{d}, the second column shows the number of abstract states for each abstraction level, and we report the accuracy, precision, and recall in the next columns.}
\label{tab:ML-Functional-fault_Cartpole}
\end{table}

\begin{table}[ht]
\centering
\color{black}\begin{tabular}{|c|c|c|c|c|}
\hline
\textit{\textbf{d}} & \begin{tabular}[c]{@{}c@{}}\textbf{Abstract} \\ \textbf{states} \end{tabular} & \textbf{Accuracy} & \textbf{Precision} & \textbf{Recall} \\ \hline \hline
0.005 & 256826 & 79\% & 84\% & 79\% \\ \hline
0.01  & 246624 & 79\% & 84\% & 79\% \\ \hline
0.05  & 195864 & 84\% & 87\% & 84\% \\ \hline
0.1   & 160142 & 88\% & 90\% & 88\% \\ \hline
0.5   & 83887  & 99\% & 99\% & 99\% \\ \hline
1     & 59169  & 99\% & 99\% & 99\% \\ \hline
5     & 19019  & 99\% & 99\% & 99\% \\ \hline
10    & 10108  & 99\% & 99\% & 99\% \\ \hline
50    & 2012   & 99\% & 99\% & 99\% \\ \hline
100   & 890    & 99\% & 99\% & 99\% \\ \hline
500   & 93     & 99\% & 99\% & 99\% \\ \hline
1000  & 38     & 99\% & 98\% & 99\% \\ \hline
5000  & 10     & 98\% & 98\% & 98\% \\ \hline
10000 & 9      & 93\% & 94\% & 93\% \\ \hline
\end{tabular}
\caption{\added{Prediction of \textit{functional faults} with \textit{Random Forest} in the \textit{Mountain Car} case study. The first column represents the abstraction level \textit{d}, the second column shows the number of abstract states for each abstraction level, and we report the accuracy, precision, and recall in the next columns.}}
\label{tab:ML-Functional-fault-MC}
\end{table}

In both case studies, we trained \textit{Random Forest} models using the previously described datasets to predict functional faults. Because of the high number of concrete states in our dataset \added{(about 250 000 in \textit{Cart-Pole} and 270 000 in \textit{Mountain Car} )}, we need to reduce the state space by using state abstraction to facilitate the learning process of the \textit{Random Forest} models. As presented in section~\ref{subsec: State Abstraction for Training}, we used the $Q^*$-irrelevance state abstraction technique~\cite{pmlr-v80-abel18a} to reduce the state space. We experimented with several values for the abstraction level \textit{d} (section~\ref{subsec:State abstraction}) and reported the prediction results in terms of precision, recall, and accuracy  \added{in Tables~\ref{tab:ML-Functional-fault_Cartpole} and~\ref{tab:ML-Functional-fault-MC}}.  
We obtained fewer abstract states when we increased \textit{d} because more concrete states were included in the same abstract states. For each value of \textit{d}, we considered a new training dataset (with different abstract states). For each dataset, we trained \textit{Random Forest} by randomly sampling 70\% of the data for training and used the remaining 30\% for testing. 

The overall prediction results for functional faults are promising. \added{As shown in Table~\ref{tab:ML-Functional-fault_Cartpole}, the best results for the prediction of functional faults yield a precision and a recall of 97\% in the \textit{Cart-Pole} problem. Similarly,  a precision and a recall of 99\% was achieved in the \textit{Mountain Car} problem as depicted in Table~\ref{tab:ML-Functional-fault-MC}.}

\added{Also, for both case studies, we observe that when we increase the state abstraction level, the accuracy of the ML classifiers improves until it plateaus and then starts to decrease as information that is essential for classification is lost.}
This highlights the importance of finding a proper state abstraction level to (1) facilitate the learning process of the ML classifiers, and (2) more accurately predict functional faults. 
\added{Note that we consider the abstraction level \textit{d} that maximizes the accuracy of the ML model while significantly decreasing the total number of distinct abstract states in the dataset. 
In the \textit{Cart-Pole} and \textit{Mountain Car} case studies, \textit{d} is equal to 1 and 500, respectively. Differences in the abstraction levels are due to the differences in the complexity of the environments and the state representations. }

\begin{tcolorbox}
\textbf{Answer to RQ2:} By using an ML classifier (based on \textit{Random Forest}) combined with state abstraction, we can accurately classify whether episodes are functional-faulty or not. Such a classifier can therefore be used as fitness functions in our search. However, finding a suitable level of state abstraction is essential to increase the learnability of the ML classifier and thus to improve the accuracy of the fault prediction results. 
\end{tcolorbox}

\subsubsection{RQ3. Can we learn accurate rules to characterize the faulty episodes of RL agents?}

We investigate the learning of interpretable rules that characterize faulty episodes to understand the conditions under which the RL agent can be expected to fail.
Consequently, we rely on interpretable ML models, namely \textit{Decision Trees}, to learn such rules. 
We assess the accuracy of decision trees and therefore our ability to learn accurate rules based on the faulty episodes that we identify with our testing approach. In practice, engineers will need to use such an approach to assess the safety of using an RL agent and targeting its retraining.

In our analysis, we consider a balanced dataset that contains (1) faulty episodes created with STARLA, and (2) non-faulty episodes obtained through random executions of the RL agent. 
We consider the same proportions of faulty and non-faulty episodes. Such a dataset would be readily available in practice to train decision trees. 
For training, we use the same type of features as for the ML model that was used to calculate one of our fitness functions (section~\ref{subsubSec:FeatureRepresnetation}). Each episode is encoded with a feature vector of binary values denoting the presence (1) or absence (0) of an abstract state in the episode. We rely on such features because the ML model that we have used to predict functional faults showed good performance using such representation. Moreover, we did not rely on the characteristics of concrete states to train the model and extract the rules due to (1) the potential complexity of state characteristics in real-world RL environments, and (2) $Q^*$-values matching abstract states are more informative since they also capture the next states of the agent and the optimal action (i.e., the agent's perception).

\begin{figure}[ht]
\centering

    \includegraphics[width=1.0\columnwidth]{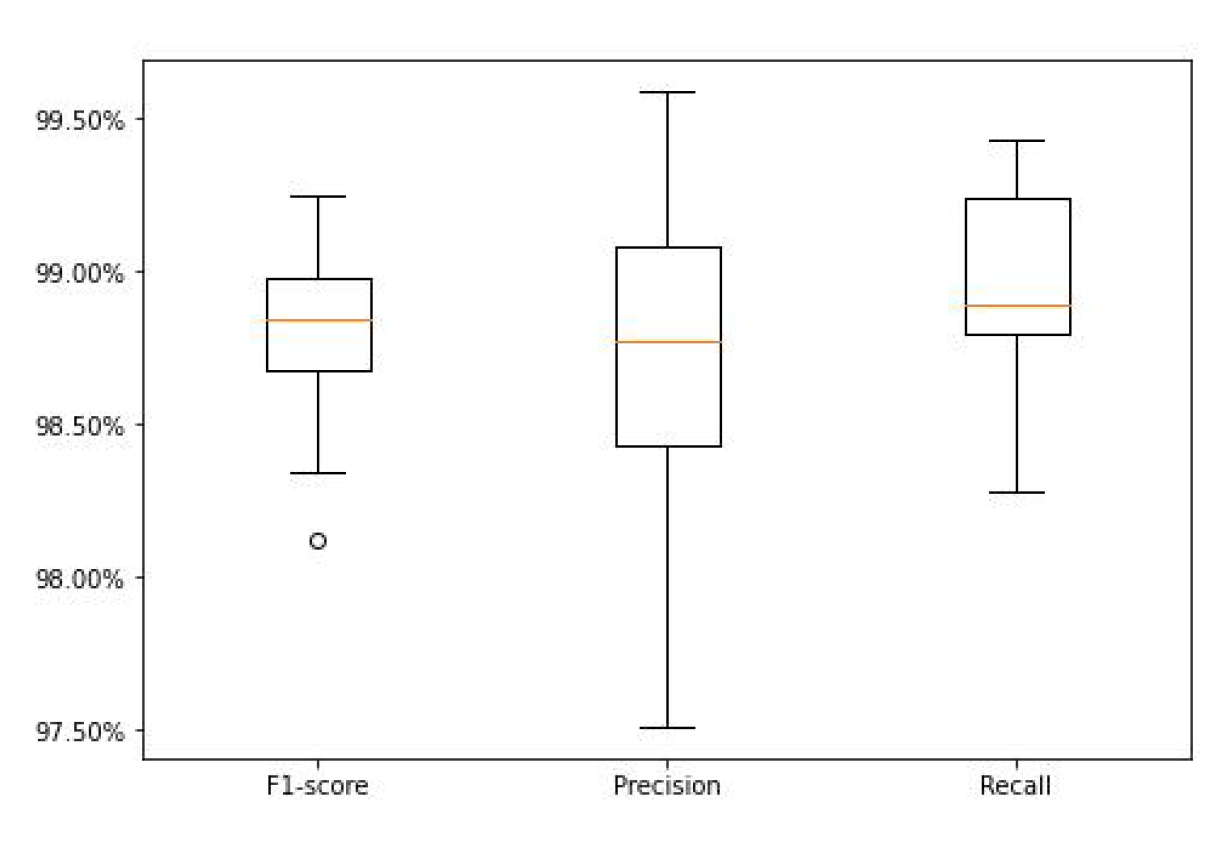}
    \caption{Accuracy of rules predicting faults in \textit{Cart-Pole}}
    \label{fig:ffault-boxplot}
\end{figure}

\begin{figure}[ht]
\centering
\includegraphics[width=1.0\columnwidth]{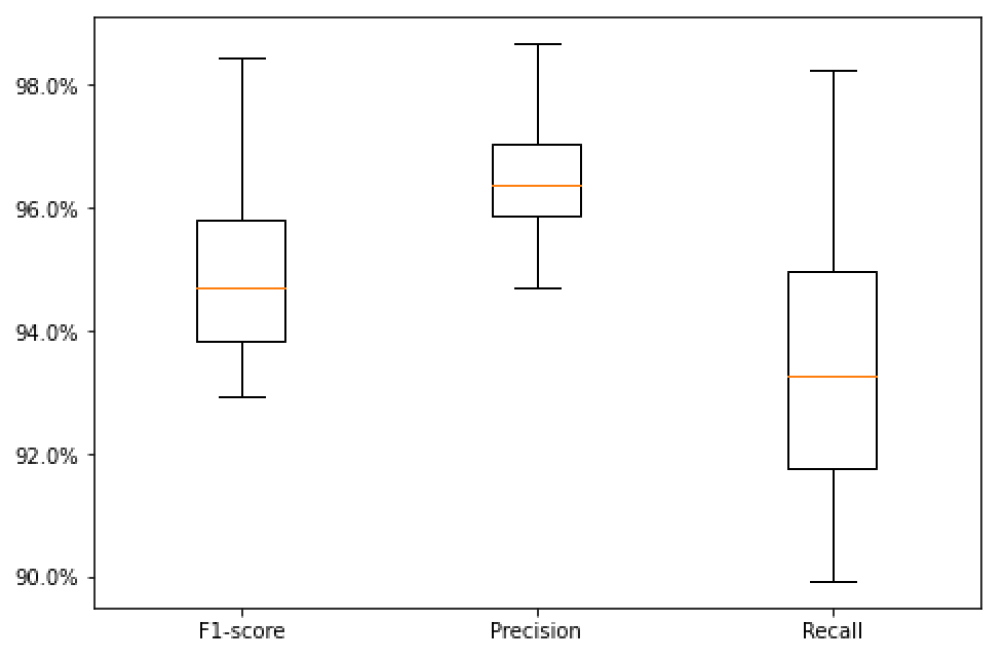}
    \caption{\added{Accuracy of rules predicting faults in \textit{Mountain Car}}}
    \label{fig:ffault-boxplot_MTC}
\end{figure}

Since we simply want to explain the faults that we detect by extracting accurate rules, we measure the accuracy of the models using K-fold cross-validation. 
We repeat the same procedure for all executions of STARLA \added{in each case study} to obtain a distribution of the accuracy of the decision trees. 
More specifically, we study the distributions of precision, recall, and F1-scores for the detected faults and report the results \added{in Figures~\ref{fig:ffault-boxplot} and~\ref{fig:ffault-boxplot_MTC}.}

\added{As shown in Figures~\ref{fig:ffault-boxplot} and~\ref{fig:ffault-boxplot_MTC}, we learned highly accurate decision trees and therefore rules (tree paths) that characterize faults in RL agents. Indeed, rules predicting faults in the \textit{Cart-Pole} case study have a median precision of 98.75\%, a recall of 98.90\%, and an F1-score of 98.85\%. }

\added{Furthermore, rules predicting functional faults in the \textit{Mountain Car} case study have a median precision of 96.25\%, a recall of 93\%, and an F1-score of 94.75\%.}  
The rules that we extract consist of conjunctions of features capturing the presence or absence of abstract states in an episode. 

We provide in the following an example of a rule that we obtained in the \textit{Cart-Pole} problem: 
\begin{align*}
\begin{split}
\textbf{R1:} \, \textit{not}(S^\phi_5) \; \textit{and} \; S^\phi_{12} \; \textit{and} \, S^\phi_{23} \\
\end{split}
\end{align*}
where rule \textit{R1} states that an episode is faulty if there are no concrete states in the episode that belong to abstract state $S^\phi_5$ and we have at least two concrete states matching abstract state $S^\phi_{12}$ and $S^\phi_{23}$, respectively.

\begin{figure}[ht]
    \centering
    \includegraphics[width=\columnwidth]{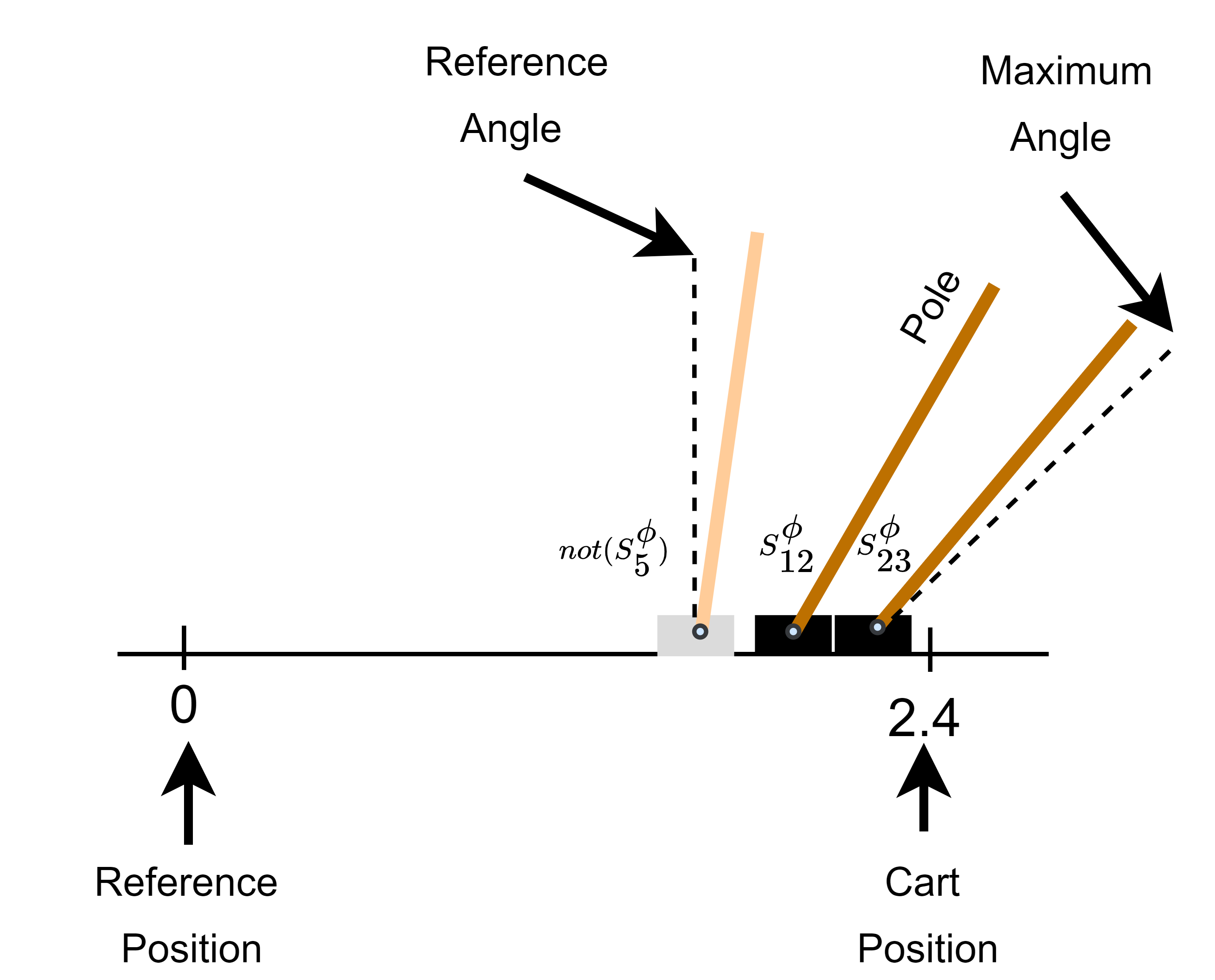}
    \caption{Interpretation of Rule \textit{R1}. Each cart represents one abstract state. The gray cart depicts the state of the system in abstract state $S^\phi_5$, which should be absent in the episode. The black carts represent the presence of abstract states $S^\phi_{12}$ and $S^\phi_{23}$, respectively. Having both latter states appearing in an episode and not having the state on the left is highly likely to lead to a fault.}
    \label{fig:RuleExample}
\end{figure}

From a practical standpoint, such highly accurate rules can help developers understand, with high confidence, the conditions under which the agent fails. For example, one can analyze the concrete states that correspond to abstract states that lead to faults to extract real-world conditions of failure. For example, to interpret the rule \textit{R1}, we first extract all faulty episodes following this rule. Then, we extract from these episodes all concrete states belonging to the abstract states that must be present according to \textit{R1} (i.e., $S^\phi_{12}$ and $S^\phi_{23}$, respectively).
For abstract states that the rule specifies as absent (abstract state $S^\phi_5$ in our example), we extract the set of all corresponding concrete states from all episodes in the final dataset.
Finally, for each abstract state in the rule, we analyze the distribution of each characteristic of the corresponding concrete states (e.g., the position of the cart in the \textit{Cart-Pole} environment, the velocity, the angle of the pole, and the angular velocity) to interpret the situations under which the agent fails. Due to space limitations, we include the box plots of the distributions of the states' characteristics in our replication package. 
Note that we did not rely directly on the abstract states' $Q^*$-values to understand the failing conditions of the agent since they are not easily interpretable. We rely on the median values of the distribution of the states' characteristics to illustrate each abstract state and hence the failing conditions. 
We illustrate these conditions in Figure~\ref{fig:RuleExample}.
Our analysis shows that the presence of abstract states $S^\phi_{12}$ and $S^\phi_{23}$ represent situations where the cart is close to the right border of the track and the pole strongly leans toward the right. To compensate for the large angle of the pole, shown in the figure, the agent has no choice but to push the cart to the right, which results in a fault because the border is crossed. Moreover, abstract state $S^\phi_{5}$ represents a situation where (1) the angle of the pole is not large, and (2) the position of the cart is toward the right but not close to the border. In this situation, the agent will be able to control the pole in the remaining area and keep the pole upright without crossing the border, which justifies why this abstract state is absent in faulty episodes that satisfy rule \textit{R1}. Note that we only provide an example of a faulty rule from the \textit{Cart-Pole} case study. Different rules that consist of more complex combinations of different abstract states can be extracted and therefore analyzed. Such interpretable rules can thus assist engineers to ensure safety and analyze risks prior to deploying the agent. 
\added{We acknowledge that the extracted rules from the detected faulty episodes are not sufficient to evaluate the risk of deploying RL agents. However, characterizing the DRL agent's faulty episodes, as we automatically do, is indeed a necessary piece of information for risk analysis. If they are accurate, such rules can be used to understand the conditions under which the agent is likely to fail.}
\begin{tcolorbox}
\textbf{Answer to RQ3:} By using our search-based technique and interpretable ML models, such as \textit{Decision Trees}, we can accurately learn interpretable rules that characterize the faulty episodes of RL agents. Such rules can then serve as the basis for risk analysis before deployment of the agent to avoid safety violations.      
\end{tcolorbox}

\section{Discussions}
\label{Sec:Discussitons}

In this paper we propose STARLA, a search-based approach to detect faulty episodes of an RL agent. To the best of our knowledge, this is the first testing approach focused on testing the agent's policy and detecting what we call functional faults.

\noindent \added{\textbf{Simulation cost.}} We rely on a small proportion of the training data of an RL agent and do not need to access the internals of the RL-based system, hence the data-box nature of our solution. Our testing approach outperforms Random Testing of RL agents because we were able to find significantly more functional faults with the same simulation budget. However, Random Testing might outperform our testing approach in simple environments with fast simulations since it typically generates a much higher number of episodes, including faulty episodes. Nonetheless, RL agents are generally used in complex environments (e.g., cyber-physical systems), where the simulation and therefore test execution costs are very high. Narrowing the search toward the faulty space of the agent is therefore crucial to optimize RL testing in a scalable way.  

\noindent \added{\textbf{Feature representation.}} Relying on state abstraction helped us to reduce the search space and increase the learnability of the ML models that we used to (1) calculate the probabilities of functional faults, and (2) extract and interpret the rules characterizing faulty episodes. However, depending on the type of the RL task, in practice, one needs to select the right state abstraction type and level to effectively guide the search toward finding faults in the RL agent. State abstraction has allowed us to extract accurate rules predicting the presence of faults and thus enables effective risk analysis. 
Although we investigated different feature representations, such as encoding episodes with sequences of abstract states, the accuracy of the ML model was only slightly improved. Therefore, it was considered not worth the additional complexity to account for such sequential information.

\noindent \added{\textbf{State abstraction.} As described in section~\ref{Subsec:Crossover}, we rely on state abstraction to find matching states for crossover points. Based on the applied abstraction method, the matched abstract states may not capture the exact physical situations. However, they correspond to concrete states leading to similar expected rewards for the same selected actions. State abstraction is one of the key heuristics to enable an effective search in our context. If we find two similar states where the agent has similar $Q^*$-values, we consider that those states are similar enough to perform the crossover. Like any heuristic, it needs to be evaluated. Therefore, to ensure the validity of the generated episodes, we validated the final created episodes by re-executing them as explained in section~\ref{sec:execute}. We also re-executed these episodes to check their consistency with the policy of the agent and whether they actually trigger failures. \minor{To further investigate the physical similarity of concrete states from the same abstract states, we analyzed the distribution of concrete state parameters (e.g., position, velocity, pole angle) in abstract states. We checked the abstract states in the rules characterizing the faulty episodes and have included examples of these distributions in our replication package. Our analysis revealed low variance in the distributions of parameters, indicating that concrete states within the same abstract state exhibit similar physical characteristics and $Q^*$-values.} Our validation also shows that the states of the environment and in the episodes where deviations occur are not frequent and have very low cosine distances. Furthermore, the results confirmed that such abstraction still leads to good search guidance since our approach outperforms Random Testing by finding significantly more functional faults.}

\noindent \added{\textbf{Functional faults.}} We should note that the number of functional faults in our case studies is relatively high since the reward function of the RL agents does not help prevent this type of fault (despite the high average reward of the agent). As mentioned in section~\ref{subsec:Case}, we have relied on standard reward functions. For example, in the \textit{Cart-Pole} environment, the agent's reward does not increase when it crosses the borders (i.e, where there is a functional fault and the episode terminates). We are using a standard, widely used, artificial benchmark to validate our testing approach, but we expect the number of functional faults in real-world RL agents to be much lower. Indeed, in more complex environments, a high penalty for functional faults could be part of the reward function to minimize them and prevent safety violations. However, as we explained in section~\ref{subsec:TestingChallenges}, this is not always enough to prevent safety-critical situations because the agent's reward could still be relatively acceptable in the presence of functional faults. \added{For example, a car may successfully reach its destination sooner by driving at a higher speed while being dangerously close to other cars. In this case, we have a high reward, because the car successfully reached the destination sooner without having any accident (the reward is defined based on the time of arrival and the penalty is applied when a collision actually occurs). However, such an episode has a high probability of functional fault since the ego car remains very close to the other cars. Relying only on the agent's reward thus makes it challenging for the search to identify functional faults. In other words, the reward and probability of functional faults are not always related. They are instead complementary and both used by STARLA to guide the search toward functional-faulty episodes. }

\added{Furthermore, we can have multiple types of functional faults related to an RL agent. In that case, we can consider the probability of each type of fault as a separate fitness function in the genetic algorithm. More specifically, we can rely on the ML model to predict the probability of each type of fault and use them as separate fitness functions. Note that considering more fitness functions to predict multiple types of functional faults is rather straightforward thanks to the use of MOSA, which is specifically designed for many-objective search problems (section~\ref{sec:Selection}).}

\noindent \added{\textbf{Initial population.}} We also investigated different sizes of the initial population (i.e., 500, 1050, and 3000) and obtained consistent results: STARLA outperformed Random Testing in terms of the total number of detected functional faults. Furthermore, we observed that the number of detected faults increased with the size of the initial population of the search. For example, in the \textit{Cart-Pole} case study, the accuracy of the faults prediction decreased for training data sets with less than 670 episodes. 
In some practical cases, costly simulations may lead to limited data for testing RL agents with STARLA. Consequently, this can affect the accuracy of ML models due to small ML training datasets and the results of the genetic search (i.e., due to the small initial population). 
From a practical standpoint, the size of the initial population is bounded by a predetermined testing budget and can be determined according to the accuracy of the ML model. Depending on the case study, we may choose the size of the initial population that maximizes the accuracy of the ML model while consuming a reasonable portion of the testing budget. 

\added{Improving the diversity in initial populations for genetic algorithms may potentially increase the quality of the search results by enabling a faster convergence to optimal solutions. As mentioned in section~\ref{subsec:initial-population}, we considered episodes starting from different initial states to diversify the initial population. In future work, we aim to investigate the use of state-of-the-art diversity metrics to guide the generation of a diverse initial population to determine whether the additional diversity computations are beneficial given that they use part of the test budget. Examples of such metrics include geometric diversity~\cite{kulesza2012determinantal}, entropy measure~\cite{albunian2020measuring}, and hamming distance~\cite{morrison2001measurement}.}

\noindent \added{\textbf{Rules characterizing faulty episodes.}} In this paper, we investigate the learning of interpretable rules that characterize faulty episodes to understand the conditions under which the RL agent can be expected to fail. If accurate, these rules can help developers to focus their analysis on specific abstract states that lead to faults, and analyze the risks related to the deployment of the RL agent. For example, after analyzing the failing rules of the agent, engineers can use abstract states leading to faults to automatically ensure safety at run-time. The agent state can be monitored to assess the risks and activate corrective measures. To prevent a failure, for example, the agent can be forced to avoid specific actions that lead to states that can violate safety requirements. 
We have relied on the presence and absence of abstract states as features to learn rules that characterize faulty episodes. However, other types of features that consider temporal information regarding the agent’s states and actions in faulty episodes (e.g., considering the sequence of abstract states) could provide additional relevant information to explain the occurrence of faulty episodes. However, learning such temporal patterns typically requires much more data to achieve accurate results and appeared to be unnecessary and therefore practically justified in our two case studies. In future work, we aim to investigate other feature representations that include temporal information to extract rules that characterize faulty episodes. We also intend to investigate in depth the use of STARLA for safety analysis before deploying RL agents. \minor{ Finally, we aim to perform a user study to understand how engineers can use such rules to make decisions about the safety of the RL agent under test.}

This paper takes a first step toward testing RL agents using a data-box genetic search. Our proposed testing approach and associated results have several practical implications.  The generated faulty episodes and the corresponding rules that characterize them can be used for safety analysis and retraining. Indeed, analyzing the states and actions in the generated faulty episodes could help developers to (1) understand the root-causes of faults in the RL agent, and (2) analyze the safety risks at run-time based on the prevalence of such root causes in practice and the consequences of identified faults. Moreover, one can retrain the agent using some of the generated faulty episodes, guided by the rules, as a mechanism to improve the policy of RL agents.

\section{Threats to Validity}
\label{Sec:Threats}
In this section, we discuss the different threats to the validity of our study and describe how we mitigated them. \\

\noindent{\textbf{Internal threats}} concern the causal relationship between the treatment and the outcome. 
Invalid episodes generated by STARLA might threaten internal validity. 
To mitigate this threat and to ensure the validity of the generated final episodes, we have relied on state abstraction and the application of realistic state transformations when using the search operators. For instance, for crossover, instead of selecting random crossover points, we have used state abstraction to find a matching pair for the crossover point. Furthermore, to ensure the validity and the exploratory aspect of the mutation operator, we alter the state of the mutation point using realistic state transformation methods to produce realistic and feasible states that could happen in the real-world environment. Finally, the validity of the episodes is checked through their execution. Thus, we only retain valid failing episodes in our final results. \minor{Frequent replacement of states during executions may pose a potential threat to the validity of our study. However, we observed that the replaced states are highly similar in terms of cosine similarity, suggesting that the physical characteristics of the replaced states are also similar.}

Our search approach relies on the specification of several thresholds that are context-dependent and vary from case to case. The threshold of the reward can change based on the expected minimum reward of the agent. For the reward fitness function in the \textit{Cart-Pole} problem, we used a threshold equal to 70, while in the \textit{Mountain Car} problem the reward threshold was -180. Based on experiments, we also realized that STARLA performs better when the threshold of the probability of functional fault fitness value is 95\% and the threshold of the certainty level is 0.04. It is important to fine-tune these parameters for each case study to get optimal detection results. \minor{We acknowledge that the fine-tuning of STARLA parameters may require additional efforts from practitioners for more complex RL problems. Therefore, in future work we plan to investigate how to fine-tune STARLA parameters for more complex problems.}

The choice of an inappropriate state abstraction method and level might also be a threat. To mitigate it, we have studied several state abstraction methods and have tried different abstraction levels to train our ML model. We have selected the best abstraction level that maximizes the accuracy of the model and significantly decreases the number of abstract states.

The current solution does not consider newly seen abstract states during testing. We acknowledge that this is a limitation of our testing approach, though any ML solution is always based on incomplete features, and what matters is whether the guidance provided to the test process is sufficiently effective. To mitigate the risk of missing abstract states in our feature representation, we have relied on a state abstraction method that considers a large number of concrete states in both the training phase and random executions of the RL agent. We computed the percentage of newly seen abstract states during the execution of the RL agent. We observed, on average, eight new abstract states out of 93 in the \textit{Mountain Car} problem and 209 new abstract states out of 1035 in the \textit{Cart-Pole} problem. Nevertheless, our results show that we trained accurate models based on known abstract states in both case studies. \minor{To further enhance the accuracy of the ML model, we can increase the size of the training dataset, thereby expanding the range of abstract states from which the model can learn. Additionally, we can retrain the ML model after newly identified abstract states are found, enabling the model to incorporate these states into the decision-making process. }

\noindent{\textbf{Conclusion threats}} are concerned with the relationship between treatment and outcome. The randomness in our search approach leads to the generation of different episodes after each run of STARLA. 
To mitigate this threat in our experiments, we executed several runs of our search method and studied the distribution of the number of the detected faults for both our method and Random Testing.

\noindent{\textbf{Reliability threats}} concern the replicability of our study results. We rely on publicly available RL environments and provide online all the materials required to replicate our study results. This includes the set of the executed and generated episodes and the different configurations that we used in our experiments.

\noindent{\textbf{External threats}} concern the generalizability of our study. Due to the high computational expense of our experiments and the lack of publicly available, realistic RL agents, we relied on two case studies in this paper. However, we mitigated this threat by using widely studied RL tasks which are considered as valid benchmark problems in many RL-related studies~\cite{Behzadan2019AdversarialEO,Behzadan2019SequentialTF,DBLP:journals/corr/abs-2006-05032,pattanaik2018robust}. However, our approach is customizable and can be applied to any other RL problem. Furthermore, in future work we aim to apply our testing approach on other RL problems to generalize the obtained results.

\section{Related Work} 
\label{Sec:RW}

Several approaches have been proposed in the literature to study the safety of RL agents during the training and execution phases. However, limited research has been targeted at testing RL-based systems.

\minor{In a very recent study, Tapple \textit{et al.}~\cite{tappler2022search} presented a search-based testing method for RL agents. The method relies on a backtracking-based, depth-first search algorithm~\cite{tarjan1972depth}that is used in the RL agent's execution to identify a reference trace that solves the RL task and contains a set of boundary states that can lead to unsafe states. Test suites are generated by extracting the action traces that lead to these boundary states from the initial state. The main objective of these test suites is therefore to guide the RL agent toward safety-critical states. Furthermore, to evaluate the agent's performance, performance test suites are generated by using a genetic algorithm to create a diverse set of action traces starting from the reference trace. The average reward gained by the policy of the agent is then compared to the average reward from the fuzz trace executions to evaluate the agent's performance. However, this testing approach is only applicable to RL agents with a stochastic policy, while our focus is on testing agents with a deterministic policy interacting with a stochastic environment (which is normally the case in safety-critical domains). Another limitation is that this approach relies on finding a reference trace that (1) solves the RL task, and (2) contains boundary states that lead to all unsafe states. Given these requirements, such a reference trace may be difficult to find and may not lead to all possible safety-critical scenarios. Indeed, while this framework can identify safety-critical situations near boundary states, it may not be able to identify all potential safety issues in more complex environments. Finally, the approach requires the repetitive execution of all possible actions from the different states in the reference trace, making it computationally intensive and highly sensitive to the size of the action space.} 

Nikanjam \textit{et al.}~\cite{nikanjam2022faults} presented a taxonomy of DRL faults and a tool to locate these faults in DRL programs. To build the taxonomy, they analyzed DRL programs on GitHub, mined Stack Overflow posts, and conducted a survey with 19 developers. They proposed \textit{DRLinter}, a model-based fault detection tool that relies on static analysis of DRL programs and graph transformation rules to locate existing faults in DRL source code. Although we have similar objectives, this work differs greatly from ours, as we detect faults related to the execution of RL agents through the search for and generation of faulty episodes. Nonetheless, this work may complement our approach and could be used as a root cause analysis mechanism of the faults reported by our search approach.

Trujillo \textit{et al.}~\cite{trujillo2020does} studied the reliability of neuron coverage~\cite{ma2018deepgauge} in testing DRL systems. They studied the correlation between coverage metrics and rewards obtained by two different models of DQN that were implemented for the \textit{Mountain Car} problem~\cite{sutton2018andrew}. They show that neuron coverage is not correlated to the agent's reward. For instance, reaching high coverage does not necessarily mean success in an RL task in terms of reward. They also showed that maximum coverage is obtained through excessive exploration of the agent, which leads to exploration of different actions that do not help maximize the agent's reward. Finally, they conclude that neuron coverage is not suitable to guide the testing of DRL systems. \minor{We therefore do not use search guided by neuron coverage as a baseline in our work.}

Several approaches have been proposed in the literature to study the robustness of RL agents against adversarial attacks~\cite{ilahi2021challenges,Huang2017AdversarialAO,pan2021improving}. For example, Ilahi \textit{et al.}~\cite{ilahi2021challenges} studied 28 adversarial attacks on RL and provided a taxonomy of existing attacks in the literature. They considered attacks that rely on perturbing (1) the state space, (2) the reward space, (3) the action space, and (4) the model space, where one can perturb the model's learned parameters. They show that although many defense approaches are proposed to increase the safety of DRL-based systems, the robustness of such systems to all possible adversarial attacks is still an open issue. This is because the proposed defense techniques in the literature can respond to the types of attacks they are built for. Besides, Moosavi-Dezfooli \textit{et al.}~\cite{DBLP:journals/corr/Moosavi-Dezfooli16} argue that regardless of the number of adversarial examples added to the training data, they were able to generate new adversarial examples to alter the normal behavior of the system.

Huang \textit{et al.}~\cite{Huang2017AdversarialAO} studied the robustness of neural network policies in the presence of adversaries. They studied the effectiveness of black-box and white-box adversarial attacks on policy networks such as DQN~\cite{mnih2013playing}, TRPO~\cite{schulman2017trust}, and A3C~\cite{mnih2016asynchronous}, trained on Atari games~\cite{2013}. They showed that adversarial attacks can significantly degrade the performance of the agent, even with small, imperceptible perturbations. 

Pan \textit{et al.}~\cite{pan2021improving} studied the robustness of the reinforcement learning agent in the specific learning task of power system control. They proposed a new adversary in both white-box and black-box (using a surrogate model) scenarios. They studied the effectiveness of their method and compared it with random and weighted adversarial attacks previously proposed for power system controls~\cite{marot2020l2rpn,omnes2021adversarial}. Moreover, they studied the robustness improvement of the agent trained with adversarial training. 

Other existing approaches in the literature~\cite{pattanaik2018robust,tan2020robustifying,DBLP:conf/iclr/KosS17,10.1145/3450267.3450537,DBLP:journals/corr/abs-2008-01825,DBLP:journals/corr/abs-2007-07176,Behzadan2018MitigationOP,behzadan2017whatever,Han2018ReinforcementLF} have proposed adversarial training techniques to increase the robustness of RL agents to adversarial attacks. 
For example, Pattanaik \textit{et al.}~\cite{pattanaik2018robust} proposed a training approach for DRL agents to increase their robustness to gradient-based adversarial attacks. They train the agent using adversarial samples generated from gradient-based attacks. They show that adding noise to the training episodes increases the robustness of the DRL agent to adversarial attacks.

Tan \textit{et al.}~\cite{tan2020robustifying} also proposed an adversarial training approach for DRL agents used for decision and control tasks. The purpose of their training approach is to increase the robustness of DRL agents against adversarial perturbations to the action space (within specific attack budgets). Consequently, they relied on gradient-based white-box adversarial attacks during the training phase of a DRL agent. They show that the proposed method increases the robustness of the agent against similar attacks.

These works differ from our testing approach, as we do not focus on the robustness of RL agents to adversarial attacks. Rather, we test the policies of RL agents, without using any of their internal information, by relying on a genetic search to effectively find faulty episodes. 

\section{Conclusion}
\label{Sec:Conclusion}

In this paper, we propose STARLA, a data-box search-based approach to test the policy of DRL agents by effectively searching for faulty episodes. We rely on a dedicated genetic algorithm to detect functional faults. We make use of an ML model to predict DRL faults and guide the search toward faulty episodes. To this end, we use state abstraction techniques to group similar states of the agent and significantly reduce the state space. This helped us to increase the learnability of the ML models and build customized search operators. We showed that STARLA outperforms Random Testing as we find significantly more faults when considering the same testing budget. 
We also investigated how to extract rules that characterize the faulty episodes of RL agents using our search results. The goal was to help developers understand the conditions under which the agent fails and thus assess the risks of deploying the RL agent. 

In future work, we aim to detect other types of faults, such as reward faults, and investigate the retraining of the RL agent using the generated faulty episodes. We aim to study the effectiveness of such episodes in improving the agent’s policy. We also want to support the safety of RL-based critical systems by providing mechanisms based on ML and state abstraction to identify sub-episodes that may lead to hazards or critical faults.

\section*{Acknowledgements}

This work was supported by a research grant from General Motors as well as the Canada Research Chair and Discovery Grant programs of the Natural Sciences and Engineering Research Council of Canada (NSERC).
We express our gratitude to Haq \textit{et al.}~\cite{Haq2021} for generously sharing the implementation of their search algorithm, which greatly facilitated our study.

\bibliographystyle{IEEEtran}
\bibliography{main.bib} 


\begin{IEEEbiography}[{\includegraphics[width=1in,height=1.25in,clip,keepaspectratio]{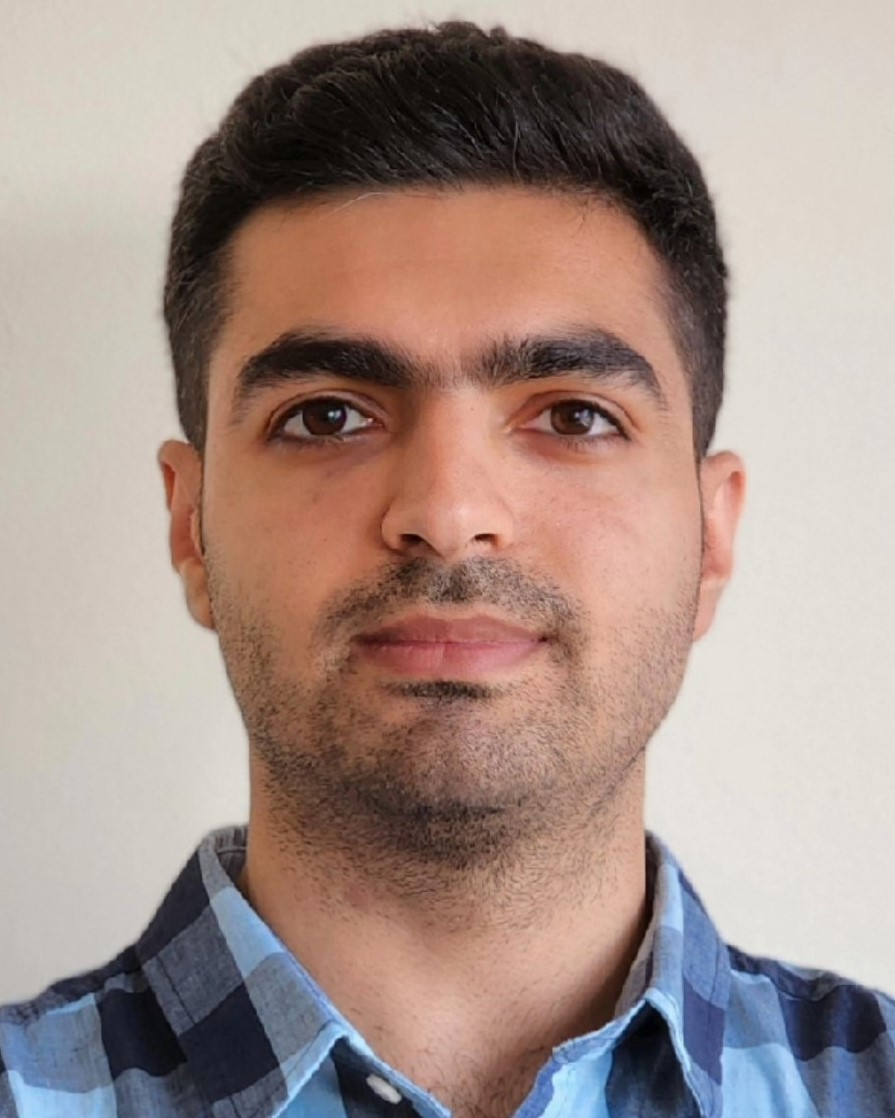}}]{Amirhossein Zolfagharian}
is a member of Nanda Lab and is currently pursuing a PhD in the School of EECS, University of Ottawa. He gained valuable practical experience from his internship at General Motors' research and development lab in the United States. Throughout his academic career, he has been the recipient of several academic awards, including a PhD admission scholarship, an international doctoral scholarship from the University of Ottawa, and an honorable award for admission to the master's program in computer science at Amirkabir University of Technology. In 2017, he was ranked as the 4th-best student among all computer science students at Amirkabir University, placing him in the top 5\% of his class in GPA. His research interests primarily focus on machine learning, empirical software engineering, and testing and verification of RL-based systems.

\end{IEEEbiography}

\begin{IEEEbiography}
[{\includegraphics[width=1in,height=1.25in,clip,keepaspectratio]{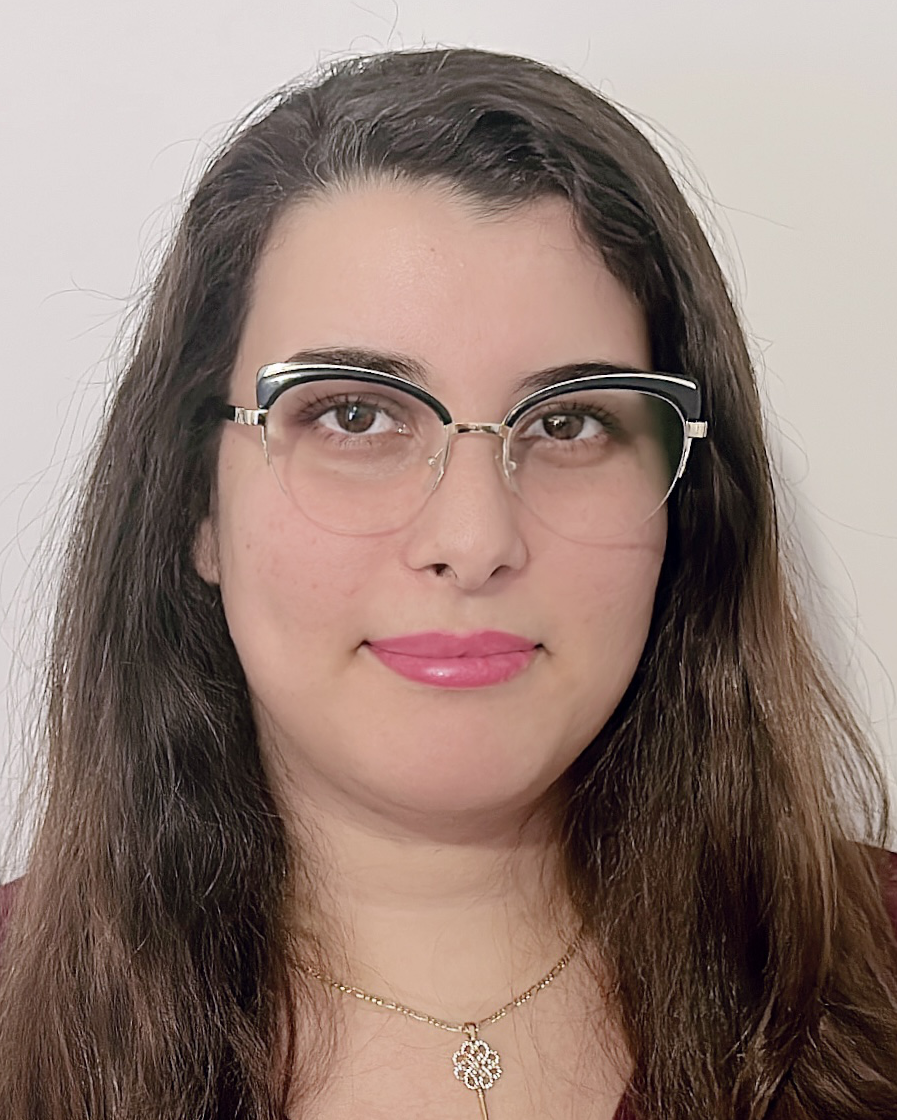}}]
    {Manel Abdellatif} received her PhD in computer science from Polytechnique Montréal, Canada (2021). She is a faculty member at École de Technologie Supérieure, Canada. She was a postdoctoral fellow at the School of EECS, University of Ottawa (2022). She received her master's degree in Information Technology from École de Technologie Supérieure (2016) and earned her bachelor's degree from École Nationale d'Ingénieurs de Tunis (2013). She served as a program committee member and a reviewer in several journals and conferences. Her research interests include testing machine learning-based systems, service computing, and empirical software engineering.
\end{IEEEbiography}

\begin{IEEEbiography}
[{\includegraphics[width=1in,height=1.25in,clip,keepaspectratio]{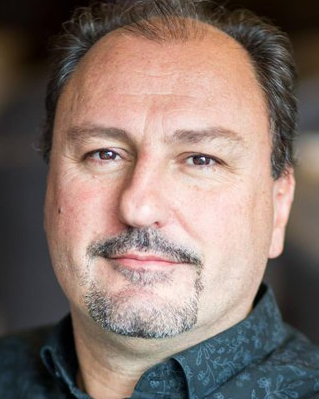}}]
    {Lionel Briand} is professor of software engineering and has shared appointments between (1) School of EECS, University of Ottawa, Canada and (2) SnT centre for Security, Reliability, and Trust, University of Luxembourg. He is the head of the SVV department at the SnT Centre and a Canada Research Chair in Intelligent Software Dependability and Compliance (Tier 1). He has conducted applied research in collaboration with industry for more than 25 years, including projects in the automotive, aerospace, manufacturing, financial, and energy domains. He is a fellow of the IEEE and ACM. He was also granted the IEEE Computer Society Harlan Mills award (2012), the IEEE Reliability Society Engineer-of-the-year award (2013), and the ACM SIGSOFT Outstanding Research Award (2022) for his work on software testing and verification. More details can be found at: http://www.lbriand.info.
\end{IEEEbiography}

\begin{IEEEbiography}
[{\includegraphics[width=1in,height=1.25in,clip,keepaspectratio]{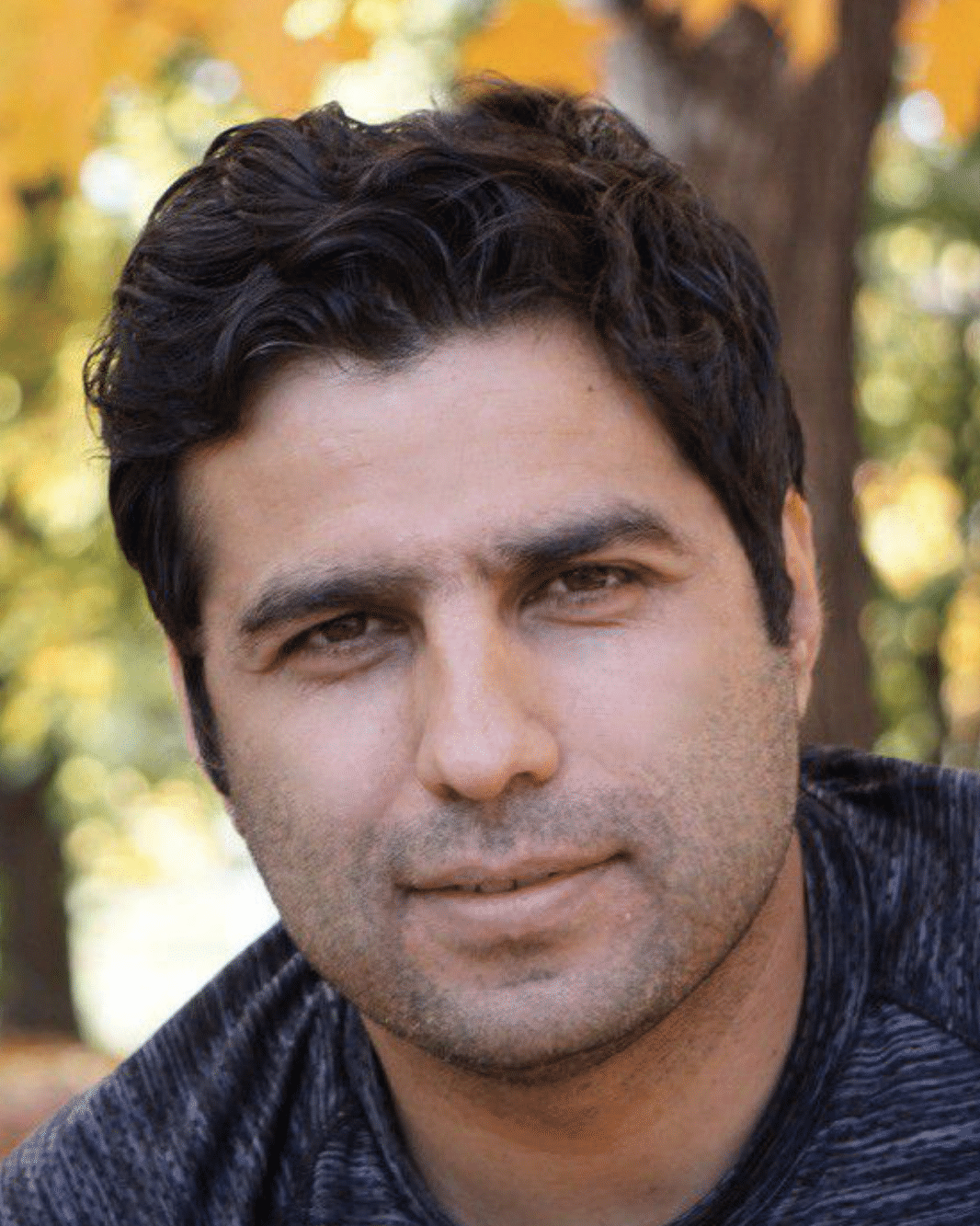}}]
    {Mojtaba Bagherzadeh} is a highly experienced Software Engineer with a proven track record of success in both industry and academia. He currently works as a software engineer at Cisco Systems and has previously worked as a software developer at IBM and a startup company. He has contributed to this research during his tenure as a postdoctoral researcher at the University of Ottawa. He obtained his PhD in Computer Science from Queen’s University, Canada, in 2019. His research interests include testing and debugging machine learning-based systems, model-driven engineering, software testing, and empirical software engineering.
\end{IEEEbiography}

\vskip 0pt plus -1fil

\begin{IEEEbiography}
[{\includegraphics[width=1in,height=1.25in,clip,keepaspectratio]{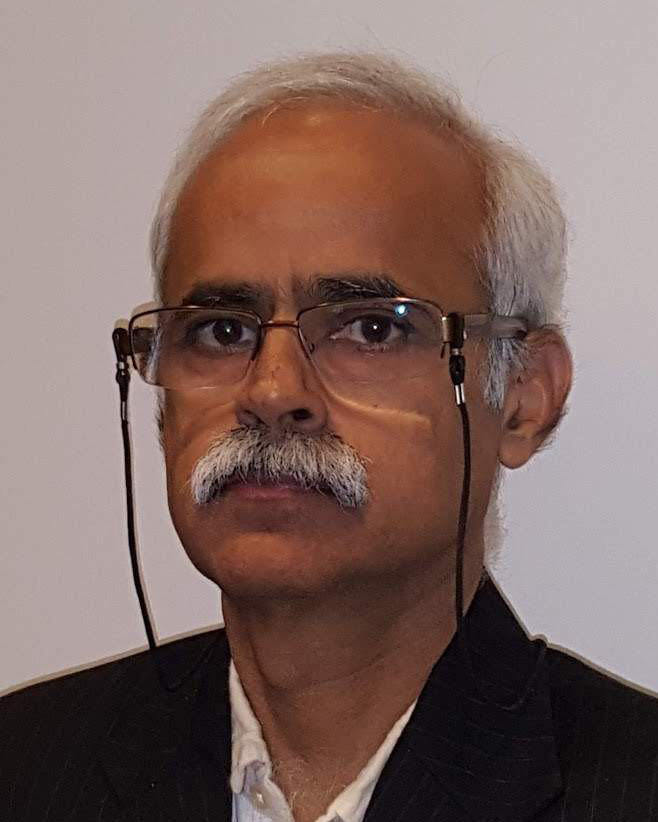}}]
    {Ramesh S}, Senior Technical Fellow,  has been with General Motors Research and Development (R\&D) for more than 15 years conducting and leading advanced research projects in the areas of model-based development of embedded systems and software, rigorous verification and validation, and more recently, AI/ML based systems. Prior to joining GM R\&D, he was a full professor in the department of Computer Science and Engineering at the Indian Institute of Technology Bombay, India, where he co-founded a Centre for Formal Design and Verification of Software. He has published more than 125 research papers in international journals and conferences and authored many patents in the areas of modeling, analysis, and verification of embedded systems and software. He has been on the program committees of several international research conferences and on the editorial boards of journals. He leads an USCAR committee and serves as an expert in ISO and SAE committees to develop guidelines for AI/ML based systems.
\end{IEEEbiography}

\end{document}